\documentclass[aps,prb,twocolumn,superscriptaddress,letterpaper]{revtex4-2}

\usepackage{graphicx}
\usepackage{bm}
\usepackage{amssymb}
\usepackage{color}
\usepackage{amsmath}
\usepackage{ulem}
\usepackage[colorlinks=true,linkcolor=blue,anchorcolor=black,citecolor=blue,filecolor=black,menucolor=black,runcolor=black,urlcolor=blue]{hyperref}

\begin{document}

\title{Suppression of Non-Hermitian Skin Effect by Pseudomagnetic Field in Honeycomb Lattice}

\author{Kai Shao}
\email{shaokai@ynu.edu.cn}
\affiliation{School of Physics and Astronomy, Yunnan Key Laboratory for Quantum Information, Yunnan University, Kunming 650091, People¡¯s Republic of China}
\affiliation{National Laboratory of Solid State Microstructures, School of Physics, and Collaborative Innovation Center of Advanced Microstructures, Nanjing University, Nanjing 210093, China}

\author{Kun Luo}
\affiliation{College of Physics and Electronic Science, Hubei Normal University, Huangshi, 435002, China}


\begin{abstract}
Magnetic suppression of the non-Hermitian skin effect (NHSE) offers a viable route for its control. While the NHSE has been realized in various classical-wave platforms, only pseudomagnetic fields (PMFs), which preserve time-reversal symmetry, can be engineered in such systems; however, their interplay with the NHSE remains underexplored. Here, we investigate this interplay in a non-Hermitian honeycomb lattice by considering two distinct mechanisms for generating PMFs: monotonically increasing strain and spatially modulated gain and loss. We show that in both scenarios, PMFs can efficiently suppress the NHSE by driving skin modes into the bulk, accompanied by a reduction of the skin topological area and a contraction of the complex-energy spectrum under periodic boundary conditions. This mechanism is insensitive to boundary details and holds for various edge terminations, including zigzag, bearded, armchair, and twig edges. Our results establish PMFs as a versatile and effective means to control the NHSE, and point toward feasible implementations in a broad range of artificial platforms, including photonic and acoustic metamaterials as well as topolectrical circuits.
\end{abstract}

\maketitle

\section{INTRODUCTION}

Non-Hermitian systems, whose Hamiltonians violate the Hermiticity condition of conventional quantum mechanics, have recently attracted significant attention owing to their unique spectral and topological characteristics absent in Hermitian counterparts~\cite{El-Ganainy2018nonNP,GongZongping2020non_Review,Kunst_RMP_2021,moiseyev2011non,Bender_1998_PRL,heiss2012physics,Unidirectional_2011_PRL,Yao2018PRL}.
Physical properties pertinent to non-Hermitian systems, such as complex eigenvalues~\cite{Bender_2002_PRL,Lee_2016_PRL,Yao2018PRL,Yao_Wang2018PRL2,Okuma2020PRL,Cai_2025_PRL_Quantum_Classical}, nonorthogonal eigenvectors~\cite{brody2013biorthogonal,curtright2007biorthogonal,Kunst_Biorth_2018_PRL,brody2016consistency}, and exceptional points~\cite{Kunst_RMP_2021,Miri_2019_Science,heiss2012physics,el2018_NP,ozdemir2019parity,Kawabata_2019_PRL_EP,hodaei2017enhanced}, can lead to a variety of unconventional phenomena.
Among them, the non-Hermitian skin effect (NHSE)~\cite{Lee_2016_PRL,Yao2018PRL,Yao_Wang2018PRL2,Kunst_Biorth_2018_PRL,Yokomizo_2019_PRL,Song_Fei_2019_PRLChiral_Damping,Liu_Second_Order_PRL_2019,Okuma2020PRL,Borgnia_2020_PRL,Zhangkai_2020_PRL_Correspondence,zhang_NC_2022_universal,Wang_Zhong_Amoeba_2024_PRX} represents one of the most striking characteristics, where an extensive number of bulk states become exponentially localized at the system boundaries under open boundary condition (OBC).
This boundary accumulation originates from the point-gap topology of complex spectra under periodic boundary condition (PBC) and is characterized by nonzero spectral winding numbers~\cite{Yao2018PRL,GongZongping_2018_PRX_Topological_Phases,Kawabata_PRX_2019_Symmetry_Topology,Yokomizo_2019_PRL,Zhangkai_2020_PRL_Correspondence,Okuma2020PRL}.
The emergence of NHSE leads to the breakdown of conventional bulk-boundary correspondence in topological physics, thus motivating the development of the non-Bloch band theory~\cite{Yao2018PRL,Yao_Wang2018PRL2,Kunst_Biorth_2018_PRL,Yokomizo_2019_PRL,Zhangkai_2020_PRL_Correspondence,Wang_Zhong_Amoeba_2024_PRX}.

The NHSE has been theoretically predicted and experimentally observed across a broad range of physical platforms, including photonic~\cite{XuePeng_2020_NP,weidemann_science_2020topological,XuePeng_2021_PRL,XuePeng_PRL_2021_Detecting,wang2021_science_generating,zhou2023_NC_observationGDSE,longhi2024_Light_incoherent}
and acoustic systems~\cite{huang2024_Nature_Review_Physics_acoustic,zhang2021acoustic_NC,wan2023Sci_Bulle_obser_Acoustic,wen2022Comu_Phy_Aco},
electric circuits~\cite{helbig2020_NP_topolectrical_circuits,LiuShuo_2021_Res_Non_Hermitian,zou2021_NC,zhang2024_SciPost},
active matter~\cite{ghatak2020_PNAS_observation},
mechanical lattices~\cite{brandenbourger2019_NC_exp_Classic,Wang_Wei_2023_PRL_ExperimentalGeometry_Dependent},
and ultracold atomic setups~\cite{liang2022_PRL_Cold,zhao2025_Nature_Ultra_cold_FermiGas}.
Its occurrence is closely related to nonreciprocal transport~\cite{Zhangkai_2020_PRL_Correspondence,Schomerus_PRRes_2020_Nonreciprocal,zhang2021acoustic_NC,zhang_NC_2022_universal,Lilinhu_2022_PRB_Direction,Genghao_2003_PRB_Nonreciprocal,Shaokai_2024_PRL_Valley,takeda2025_arxiv_reciprocal,LuMing_SunQF_2024_PRB_unidirectional,ZhangYingqiu_2025_Laser_Photonics},
delocalization transitions~\cite{ChenShu_2019PRB_Disorder_Trans,Hughes_2021_PRB_NHDIsorder,LinQuan_2022_NC,Longhi_2019_PRL_Transition,LiuYanxia_2021_PRBPLocalization_transition,LiuYanxia_2021_PRB_Exact_mobility,Lilinhu_2021_ComPhy_Impurity,Kawabata_2023_PRX_Entanglement_Phase_Transition,XuePeng_2022_PRL_Phase_Transitions,Chakrabarty_2023_PRB_Skin_effect_delocalization,Molignini_2023_PRRes_disordered,JinWeiWu_2025_PRL_Anderson_Delocalization},
and has stimulated proposals for ultrasensitive sensing applications~\cite{Budich_2020_PRL_Sensors,Koch_2022_PRRes_sensors,Wiersig_2020_Optica_sensors,mcdonald_NC_2020_sensing}.
Moreover, recent studies have revealed that both electric fields~\cite{YiPeng_2022_PRB_Manipulating} and magnetic fields~\cite{LuMing_2021_PRL_Magnetic_Suppression,Shaokai_2022_PRB_Cyclotron,teo2024_SciBul_pseudomagnetic,XuePeng_2023_NC_Manipulating,HeGao_APR_2024_Controlling,ChaoXu_2025_PRB_Controllable} can effectively suppress the NHSE, driving the skin modes from the boundaries into the bulk and thus providing promising strategies for its control.

\begin{figure*}
	\centering
	\includegraphics[width=2\columnwidth]{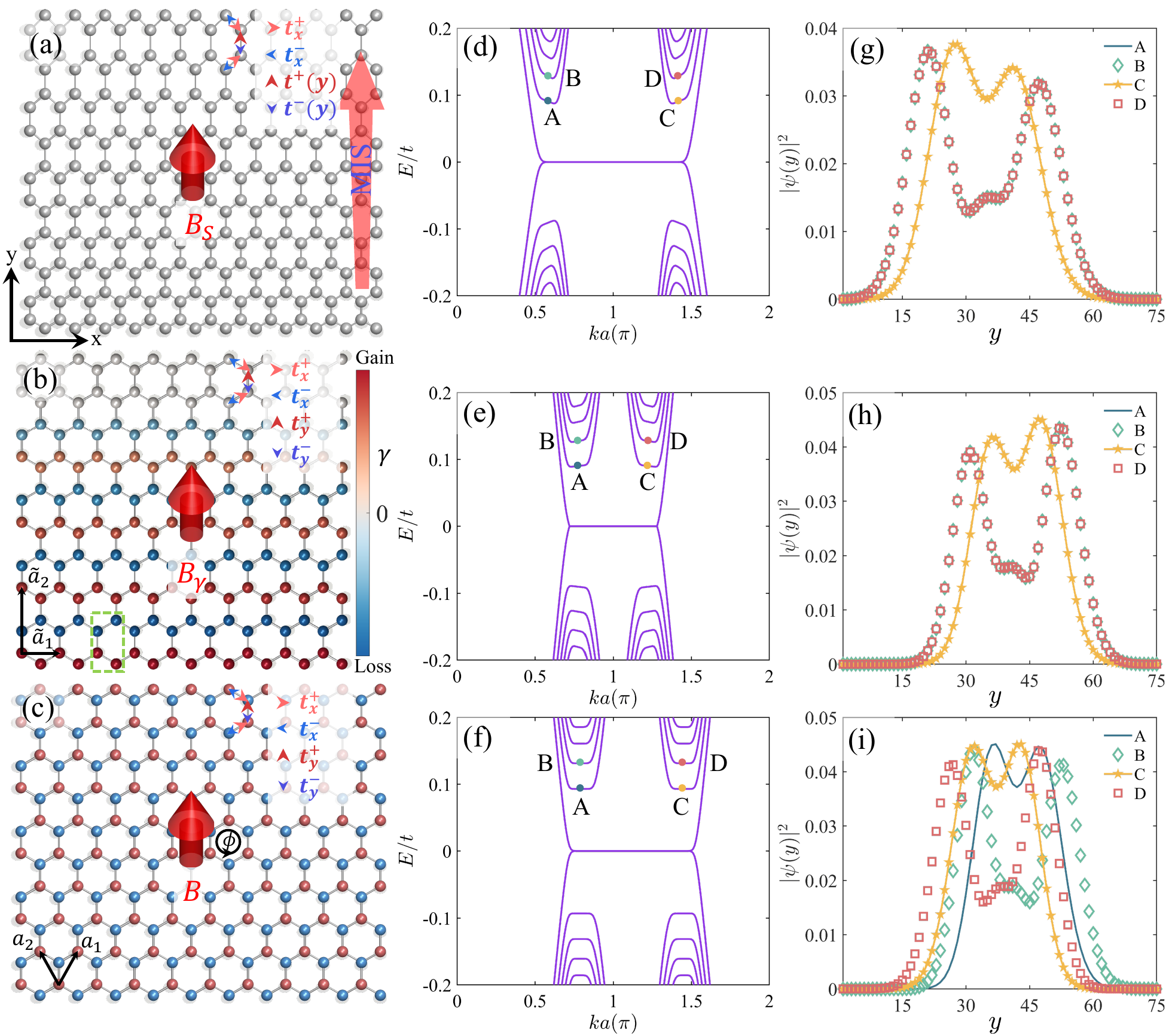}
    \caption{
    (a) Realization of a uniform PMF $B_s$ in a strained honeycomb lattice by modulating the hopping strength $t_y$ along the $y$ axis.
    (b) Generation of a PMF via spatially modulated gain and loss strength $\gamma$.
    The black arrows indicate the primitive vectors $\mathbf{\tilde{a}}_1$ and $\mathbf{\tilde{a}}_2$, and the green dashed box denotes the unit cell.
    (c) Honeycomb lattice under a real magnetic field with magnetic flux $\phi$ per plaquette.
    Nonreciprocal hoppings $\delta_x$ and $\delta_y$ can be superimposed along the $x$ and $y$ directions in (a)--(c).
    (d--f) Energy spectra corresponding to (a)--(c) with $\delta_x=\delta_y=0$.
    (g--i) Spatial distributions of the wavefunction intensity $|\psi(y)|^2$ at representative $k_x$ points marked in (d)--(f).
    Parameters: $N_y=300$, $t=1$, $t'=1.5$ in (d,g), $\beta=0.005$ in (e,h), and $\phi=\beta/\pi$ in (f,i).
    }
	\label{fig1}
\end{figure*}

Unlike real magnetic fields, which break time-reversal symmetry, pseudomagnetic
fields (PMFs) have recently emerged as a powerful approach for controlling wave localization through Landau quantization while preserving time-reversal symmetry~\cite{guinea_NP_2010_Strain,levy_Science_2010_strain}.
This scheme has been applied to a variety of classical wave platforms, enabling magnetic-field effects and the formation of bulk Landau modes in time-reversal-invariant settings~\cite{rechtsman_Np_2013_strain,ZhangBaile_2017_PRL_StrainInduced,Abbaszadeh_2017_PRL_Sonic_Landau,wen_NP_2019acousticLandau,jamadi_LightSA_2020_Direct,WangWenhui_2020_PRL_MoireFringe,XueHaoran_2020_PRL_Non_DiracCone,YanMouDengweiying_2021_PRL_Pseudomagnetic,SongWange_2022_PRL_DispersionlessCoupling,huang_2022_Nanophotonics_pattern,HeLin_2022_PRL_MagneticFieldTunable,jia_2023_Light_Sci_experimental,FanShaohui_2023_PRL_Artificial_NonAbelian,duan_2023_APL_synthetic,Barsukova_2024_NPho_direct,barczyk_2024_NPho_observation,Zhao_Wen_2024_PRL_LandauRainbow,MoQingyang_2025_PRL_Observation,liu_2025_APLPh_coupled,ZhangXiao_2025_PRL_PhotonicFlatLandau}.
Since platforms where the NHSE has been observed are typically insensitive to real magnetic fields~\cite{XuePeng_2020_NP,weidemann_science_2020topological,XuePeng_2021_PRL,XuePeng_PRL_2021_Detecting,wang2021_science_generating,zhou2023_NC_observationGDSE,longhi2024_Light_incoherent,huang2024_Nature_Review_Physics_acoustic,zhang2021acoustic_NC,wan2023Sci_Bulle_obser_Acoustic,wen2022Comu_Phy_Aco,huang2024_Nature_Review_Physics_acoustic,zhang2021acoustic_NC,wan2023Sci_Bulle_obser_Acoustic,wen2022Comu_Phy_Aco,helbig2020_NP_topolectrical_circuits,liu_2021_Research,zou2021_NC,zhang2024_SciPost,ghatak2020_PNAS_observation,brandenbourger2019_NC_exp_Classic,Wang_Wei_2023_PRL_ExperimentalGeometry_Dependent},
understanding the impact of PMFs on the NHSE is crucial for its control.
In this work, we investigate the interplay between the NHSE and PMFs on a honeycomb lattice as shown in Fig.~\ref{fig1}.
The NHSE in our model is generated by nonreciprocal hoppings;
unlike previous studies on two-dimensional square lattice system where PMFs were produced by inhomogeneous hoppings~\cite{teo2024_SciBul_pseudomagnetic}, here the PMF can be implemented either via spatial modulation of the hopping amplitudes or via spatially nonuniform gain and loss.
The orientation of the nonreciprocal hoppings controls whether the skin accumulation occurs along the $x$- or $y$-direction.
By examining spatial profiles of skin modes, the skin topological area, and the spectral area in the complex energy plane, we show that a PMF can effectively suppress the NHSE ¡ª demonstrating that a PMF can play the role of a real magnetic field in controlling the NHSE.
Finally, we discuss realistic mappings of the model to photonic, acoustic, and electric circuit platforms, which provide a practical route to manipulate wave localization in time-reversal-symmetric classical wave devices.

The paper is organized as follows. In Sec.~\ref{Sec2}, we introduce two distinct mechanisms for generating PMFs in the honeycomb lattice.
Sec.~\ref{Sec3} examines the interplay between the PMF and the NHSE, including the suppression of skin modes along both $x$ and $y$ directions.
In Sec.~\ref{Sec4}, we extend our analysis to honeycomb nanoribbons with different edge terminations¡ª bearded, armchair, and twig edges¡ªto demonstrate the robustness of PMF-induced NHSE suppression.
Finally, Sec.~\ref{Sec5} summarizes our main findings and provides an outlook for future research.

\section{Generation of the Pseudomagnetic Field}
\label{Sec2}

In this section, we introduce two complementary schemes for generating a uniform PMF in honeycomb lattices with zigzag edges.
The first scheme is Hermitian in nature and is based on a monotonic spatial modulation of the nearest-neighbor hopping amplitudes, which can be realized, for example, via nonuniform strain.
The second scheme relies on a purely non-Hermitian mechanism, where a spatially varying gain and loss profile induces an effective pseudogauge field through the dependence of Dirac point positions on the gain¨Closs strength.
Despite their distinct physical origins, both approaches lead to an effective vector potential that couples to low-energy Dirac fermions in the same manner as a real magnetic field, thereby generating a uniform PMF.

\subsection{PMF Induced by Nonuniform Strain}
\label{Sec2_1}

We consider a nearest-neighbor tight-binding model on a honeycomb lattice~\cite{Lantagne_2020_PRB_Dispersive_Landau},
\begin{equation}
H_0=\sum_i \varepsilon_i c_i^\dagger c_i
+\sum_{\langle i,j\rangle} t_{ij} c_i^\dagger c_j ,
\label{hq_Bs_dy}
\end{equation}
where $\varepsilon_i$ denotes the on-site energy and $c_i^\dagger$ ($c_i$) creates (annihilates) a particle at site $i$. To generate a PMF, we apply a monotonic increasing strain (MIS) along the $y$ direction
[Fig.~\ref{fig1}(a)], such that only the nearest-neighbor hopping amplitude along $y$ varies spatially,
while all other hoppings remain fixed.
The hopping modulation is chosen to vary linearly across the sample,
which breaks inversion symmetry and is therefore capable of generating a finite PMF~\cite{guinea_NP_2010_Strain}.

Although the MIS breaks exact translational invariance along the $y$ direction,
the hopping profile varies smoothly on a length scale much larger than the lattice constant.
In the long-wavelength limit, the system can thus be treated as locally translationally invariant,
allowing a momentum-space description within a locally periodic approximation,
the Hamiltonian of strained honeycomb lattice in
momentum space takes the form
\begin{equation}\label{Hk0}
\begin{split}
H_k=
\begin{pmatrix}
0 & t_y + t e^{i\mathbf{k}\cdot\mathbf{a}_1} + t e^{i\mathbf{k}\cdot\mathbf{a}_2} \\
t_y + t e^{-i\mathbf{k}\cdot\mathbf{a}_1} + t e^{-i\mathbf{k}\cdot\mathbf{a}_2} & 0
\end{pmatrix},
\end{split}
\end{equation}
where $t_y=t$ in the absence of strain and the primitive
vectors are given by
$\mathbf{a}_1=\left(\frac{ a}{2},\frac{\sqrt{3}a}{2}\right),
\mathbf{a}_2=\left(-\frac{a}{2},\frac{\sqrt{3}a}{2}\right),
$ and $a$ is the lattice constant.

For the pristine lattice, the conduction and valence bands touch at the Dirac points $\mathbf{K}_D$ and $\mathbf{K}'_D$. Upon introducing a position-dependent hopping amplitude $t_y = t + \frac{t' - t}{N_y}y$, where $t'$ denotes the hopping amplitude at the upper boundary and $N_y$ is the number of lattice sites along the $y$ direction, the band-touching condition necessitates a shift of the Dirac points to $\mathbf{K}_S = \mathbf{K}_D + \boldsymbol{\Delta}$. To leading order in the strain strength, this shift occurs along the $k_x$ direction and is given by $\Delta_x = \frac{2\alpha}{\sqrt{3} a} y$, with $\alpha = (t' - t) / (t N_y)$.
This expression is obtained by linearizing the tight-binding Hamiltonian around the Dirac points (see Appendix~\ref{AppA}).

The position-dependent displacement of the Dirac points can be interpreted as an effective valley-dependent
pseudo-gauge field~\cite{guinea_NP_2010_Strain,Lantagne_2020_PRB_Dispersive_Landau,HeLin_2013_PRB_PMF,FangJingYun_2024_PRB_PMF,SunJunsong_2023_PRB_PseudomagneticSquare,WuBingLan_2021_CPB_PMF}.
Adopting the conventional electromagnetic notation, the associated PMF
$\mathbf{B}_s=\nabla\times\mathbf{A}_s$ is uniform and given by
\begin{equation}
B_s=\frac{2\hbar\alpha}{\sqrt{3}ea}.
\end{equation}
As required by time-reversal symmetry, the PMF has opposite signs in the two valleys.

Expanding the Hamiltonian around the $\mathbf{K}$ and $\mathbf{K}'$ valleys yields the low-energy Dirac Hamiltonian
\begin{equation}\label{G_h_q}
h(\mathbf{q})=\hbar v_F \left[\tau  (q_x-B_s y)\sigma_x+ q_y\sigma_y\right],
\end{equation}
where $v_F$ is the Fermi velocity,
$\tau=\pm1$ labels the valley index,
$\sigma_{x,y}$ are Pauli matrices acting on the sublattice degree of freedom,
and $\mathbf{q}$ denotes the momentum measured from the Dirac point.
In this effective description, we retain the leading effect of strain through the pseudo-gauge field,
while neglecting the strain-induced anisotropy of the Fermi velocities,
which constitutes a subleading correction [cf. Appendix~\ref{AppA}].

Upon applying strain along the $y$-direction,
translational symmetry is broken in that direction while remains
preserved along $x$, allowing the band structure to
be computed. The resulting band structure under open boundary conditions along $y$ is shown in Fig.~\ref{fig1}(d).
The PMF gives rise to dispersive pseudo-Landau levels, whose slopes are opposite in the two valleys due to the inhomogeneous Fermi velocities~\cite{Lantagne_2020_PRB_Dispersive_Landau,HeLin_2013_PRB_PMF,FangJingYun_2024_PRB_PMF,SunJunsong_2023_PRB_PseudomagneticSquare,WuBingLan_2021_CPB_PMF} and time-reversal symmetry.

For comparison, we consider the effect of a real magnetic field $\mathbf{B}$ [see Fig.~\ref{fig1}(c)]. The field is incorporated via the Peierls phase $\phi_{ij} = \int_{\textbf{i}}^{\textbf{j}} \mathbf{A} \cdot d\mathbf{l} / \phi_0$ in the hopping elements, where $\mathbf{A} = (-By, 0, 0)$ and $\phi_0 = \hbar/e$ is the magnetic flux quantum. The corresponding band structure for a pristine honeycomb lattice is shown in Fig.~\ref{fig1}(f). In the low-energy regime, well-defined $n$th Landau levels appear at energies $E_n \propto \text{sgn}(n)\sqrt{B|n|}$ ($n = 0, \pm1, \pm2, \dots$). In contrast to the PMF case, these Landau levels remain perfectly flat.

To further characterize the wave functions, we examine the spatial profiles of the wave functions along the $y$ direction.
Specifically, we calculate the probability distribution $|\psi(y)|^2$ by summing the wave function intensities over the four sites within a unit cell [indicated by the green dashed box in Fig.~\ref{fig1}(b)] at selected $k_x$ points. Representative results are shown in Fig.~\ref{fig1}(g--i) for points A--D [marked in Figs.~\ref{fig1}(d--f)], corresponding to the first and second Landau levels in different valleys. In both the PMF and real MF cases, the wave functions are localized within the bulk, demonstrating  the confining effect of the gauge fields on the particles.

\subsection{PMF Induced by Spatially Modulated Gain and Loss}
\label{Sec2_2}

We now introduce a distinct, purely non-Hermitian mechanism for generating a PMF, which does not rely on lattice deformation or strain.
Instead, the effective pseudogauge field arises from the dependence of Dirac point positions on the gain¨Closs strength $(\pm i\gamma)$.
The lattice configuration is shown in Fig.~\ref{fig1}(b), where each unit cell contains four sublattices.
The momentum-space Hamiltonian reads
\begin{equation}
h_k=
\begin{pmatrix}
i\gamma & t(1+e^{-ik_x a}) & 0 & t e^{-i\sqrt{3}k_y a} \\
t(1+e^{ik_x a}) & i\gamma & t & 0 \\
0 & t & -i\gamma & t(1+e^{ik_x a}) \\
t e^{i\sqrt{3}k_y a} & 0 & t(1+e^{-ik_x a}) & -i\gamma
\end{pmatrix},
\label{hk_gain_loss}
\end{equation}
whose eigenenergies are $\{\pm E_1(k),\pm E_2(k)\}$, where
\begin{widetext}
\begin{equation}
\begin{split}
E_{1,2}(k)=
\sqrt{3t^2 + 2t^2 \cos(k_x a )- \gamma^2
\pm 2\sqrt{2}t\cos\left(\frac{k_x a }{2}\right)
\sqrt{ t^2 - 2\gamma^2 +  t^2\cos(\sqrt{3}k_y a )}}.
\end{split}
\label{Ek_gain_loss}
\end{equation}
\end{widetext}

When $\gamma=0$, the Dirac points can be determined from the condition $E_2(k)=0$, yielding $K_D=\frac{2\pi}{3a}(\pm1,0)$.
For $|\gamma|<t$, the energy spectrum remains real within the region $|k_y|<\frac{\cos^{-1}(2\gamma^2/t^2-1)}{\sqrt{3}a}$, as well as along the lines $k_x a=\pm \pi$~\cite{XueHaoran_2020_PRL_Non_DiracCone}.
The zero-energy band degeneracy points are then located at
\begin{equation}
\begin{split}
\textbf{K}^{\pm} =
\left(\begin{array}{c} \tau\theta \\ 0 \end{array}\right),
\quad \text{where } \left\{\begin{array}{l}
\tau = \pm 1, \\
\cos{\frac{\theta a}{2}} = \frac{\sqrt{t^2-\gamma^{2}}}{2t}.
\end{array}\right.
\end{split}
\label{Kd_gain_loss}
\end{equation}
Here, $\tau$ denotes which degeneracy point is being considered, analogous to the valley index in a pristine honeycomb lattice.
Because the position of the Dirac point now depends on $\gamma$, a spatial gradient in $\gamma$ effectively generates a pseudogauge field.
For instance, setting $\theta = \theta_0 + \beta y$ with $\theta_0=\frac{2\pi}{3a}$ leads to a spatial dependence of
$\gamma(y) = t \sqrt{1 - 4 \cos^{2}\left( \frac{\pi}{3} + \frac{a\beta}{2} y \right)}$.
The corresponding $\gamma$-induced pseudogauge field at $\textbf{K}^+$ is $\mathbf{A_{\gamma}}=(A_x,0)$ with $A_x=\frac{\hbar}{e}\beta y$, giving rise to a uniform perpendicular PMF $\mathbf{B_{\gamma}}=\nabla \times \mathbf{A_{\gamma}}=(0,0,B_{\gamma})$, where $B_{\gamma}=\frac{\hbar}{e}\beta$.
Since the spatially varying gain and loss are applied along the $y$ direction [see Fig.~\ref{fig1}(b)], the system still retains translational invariance along the $x$ axis, analogous to the case discussed in Sec.~\ref{Sec2_1}, the corresponding band structure is shown in Fig.~\ref{fig1}(b).
Notably, similar to the Landau levels induced by the MIS,
the system exhibits dispersive pseudo-Landau levels, as shown in Fig.~\ref{fig1}(e).
Furthermore, the spatial distributions of the wavefunctions [Fig.~\ref{fig1}(h)] confirm that the gain¨Closs-induced PMF also leads to bulk localization of eigenstates.

These results demonstrate that spatially modulated gain and loss provide an alternative, non-Hermitian route to engineer PMFs, thereby significantly extending the range of physical platforms in which PMF-controlled NHSE can be realized.

\section{Competition between the PMF and the NHSE}
\label{Sec3}

In this section, we investigate the competition between the PMF and the NHSE.
Before introducing the PMF, we first characterize the NHSE in the absence of any pseudogauge field.
This allows us to establish a clear baseline for identifying and quantifying the suppression of the NHSE once the PMF is switched on.

\begin{figure}
	\centering
	\includegraphics[width=1\columnwidth]{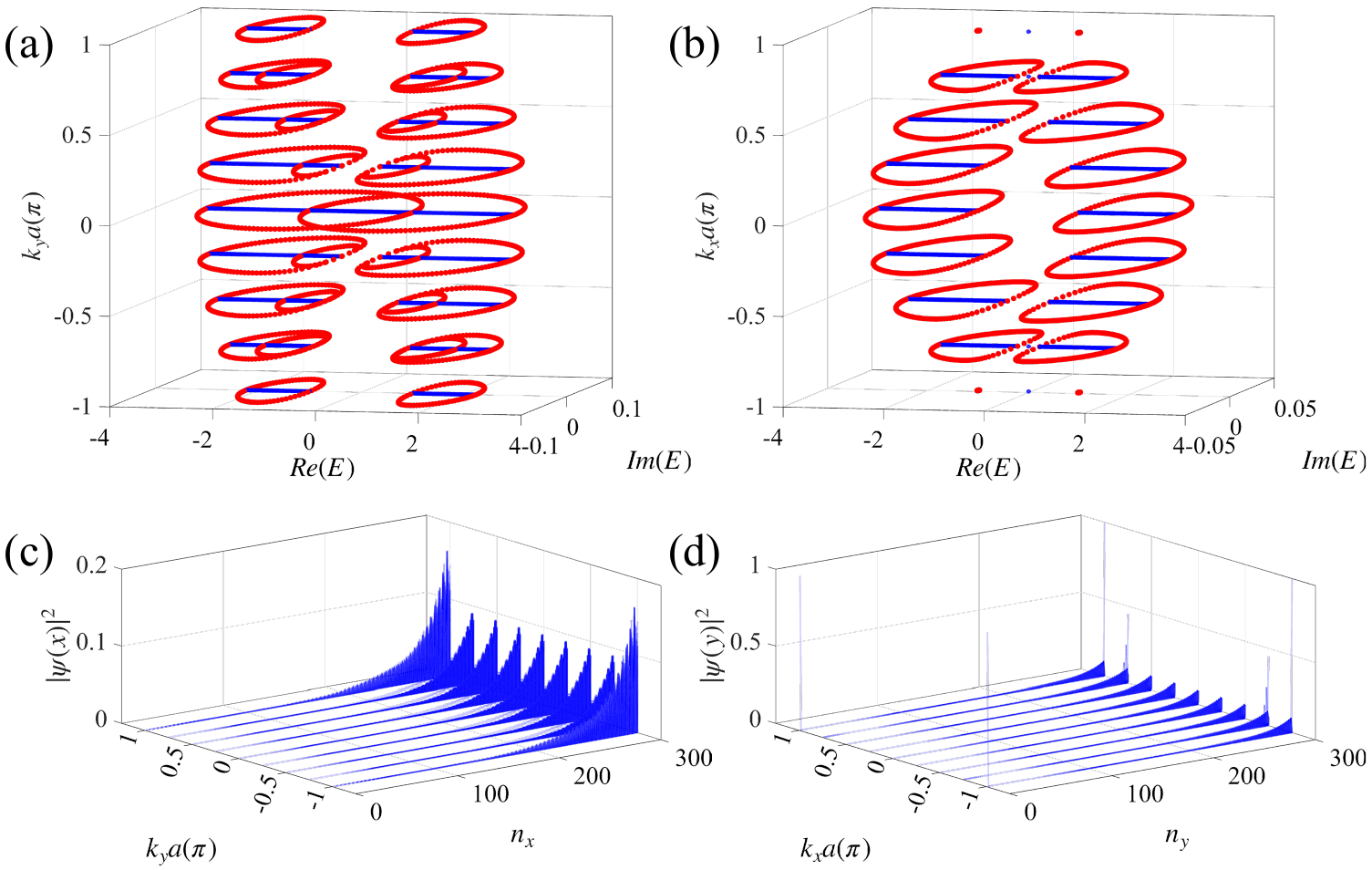}
    \caption{
    (a,b) Complex energy spectra under different boundary conditions:
    (a) $x$-PBC (red) and $x$-OBC (blue) with $\delta_x=0.05$ and $\delta_y=0$;
    (b) $y$-PBC (red) and $y$-OBC (blue) with $\delta_x=0$ and $\delta_y=0.05$.
    (c,d) Spatial distributions $|\psi(x)|^2$ and $|\psi(y)|^2$ corresponding to the blue OBC eigenenergies in (a) and (b), respectively, for a finite lattice with $N_x (N_y)=300$.
    }
	\label{fig2}
\end{figure}

\subsection{Non-Hermitian Skin Effect in the absence of a PMF}
\label{Sec3_1}

We begin by considering the NHSE without any PMF.
Non-Hermiticity is introduced through nonreciprocal nearest-neighbor hoppings,
as illustrated by the colored arrows in Fig.~\ref{fig1}(a,b).
Specifically, the hopping amplitudes along the $x$ and $y$ directions are
$t_x^{\pm}=t\pm\delta_x$ and $t_y^{\pm}=t\pm\delta_y$, respectively,
where $\delta_x$ and $\delta_y$ control the strength and direction of nonreciprocity.

To diagnose the NHSE, we compare the complex energy spectra under periodic and open boundary conditions.
Hereafter, periodic and open boundary conditions along the $x$ ($y$) direction are denoted as $x(y)$-PBC and $x(y)$-OBC, respectively.

We first induce the NHSE along the $x$ direction by choosing $\delta_x=0.05$ and $\delta_y=0$.
Under full PBC, the complex energy spectrum forms a closed loop in the complex-energy plane as $k_x a$ varies from $-\pi$ to $\pi$ for each fixed $k_y$ [red points in Fig.~\ref{fig2}(a)].
The topology of this loop is characterized by the spectral winding number
\begin{equation}
\label{Winding_num}
w(E_{0}) = \frac{1}{2\pi\mathrm{i}}
\int_{-\pi}^{\pi}
\mathrm{d}k_{x}
\frac{\mathrm{d}}{\mathrm{d}k_{x}}
\log\operatorname{det} \left[ \mathcal{H}(k_{x}) - E_{0} \right],
\end{equation}
where $E_0$ is an arbitrary reference energy inside the loop.
A nonzero winding number indicates a nontrivial point-gap topology and predicts the emergence of the NHSE upon switching to OBC~\cite{GongZongping_2018_PRX_Topological_Phases,Okuma2020PRL,Zhangkai_2020_PRL_Correspondence,Borgnia_2020_PRL}.

This can be confirmed by the spectrum under $x$-OBC, shown as blue points in Fig.~\ref{fig2}(a).
The eigenvalues collapse from the PBC loop into open arcs located inside the loop, signaling the formation of skin modes.
For each fixed $k_y$, the two-dimensional system effectively reduces to a one-dimensional non-Hermitian model along $x$ that exhibits the NHSE.
The corresponding real-space wavefunctions in a finite lattice reveal a strong accumulation of probability density at the right boundary [Fig.~\ref{fig2}(c)], providing direct evidence of the skin effect.

An analogous NHSE can be induced along the $y$ direction by choosing $\delta_x=0$ and $\delta_y=0.05$.
In this case, for each fixed $k_x$, the PBC spectrum forms a closed loop in the complex-energy plane [Fig.~\ref{fig2}(b)].
Upon imposing $y$-OBC, the spectrum collapses into open arcs, and the associated eigenstates become localized near the upper boundary of the system, as shown in Fig.~\ref{fig2}(d).
These results establish the existence of directional NHSEs controlled independently by $\delta_x$ and $\delta_y$.

It is instructive to contrast the spatial structure of the skin modes with that of the PMF-induced Landau modes discussed in Sec.~\ref{Sec2}.
As shown in Fig.~\ref{fig1}(g,h), the pseudo-Landau states generated by the PMF are localized in the bulk,
whereas the skin modes in Fig.~\ref{fig2}(c,d) accumulate exponentially at the system boundaries.
In the present model, these two localization mechanisms originate from distinct physical sources and are controlled by independent parameters:
the PMF strength ($\mathbf{B}_s$ or $\mathbf{B}_\gamma$) and the nonreciprocal hopping amplitudes ($\delta_x,\delta_y$), respectively.

We now turn on both ingredients simultaneously to explore their interplay.
Since the PMF breaks translational symmetry along the $y$ direction while preserving it along $x$,
$k_x$ remains a good quantum number whereas $k_y$ does not.
As a result, different diagnostic tools are required to analyze the suppression of the NHSE along the $x$ and $y$ directions.

\subsection{Pseudomagnetic suppression of the NHSE along the $x$ direction}
\label{Sec3_2}

We now investigate how a PMF suppresses the NHSE when the skin accumulation occurs along the $x$ direction.
Since the PMF preserves translational symmetry along $x$ while breaking it along $y$,
the spectral winding number defined in Eq.~\eqref{Winding_num} remains a suitable diagnostic for characterizing the NHSE along $x$.

We consider a semi-infinite ribbon with $x$-PBC and $y$-OBC, consisting of $N_y=300$ lattice sites along the $y$ direction.
In this configuration, the system can be regarded as an effective one-dimensional lattice with a large supercell.
For each reference energy $E_0$ in the complex plane, a nonzero winding number indicates the presence of skin modes localized along the $x$ direction under OBC.
Throughout this subsection, the nonreciprocal hopping parameters are fixed at $\delta_x=0.1$ and $\delta_y=0$,
while the strength of the PMF¡ªinduced either by MIS or by spatially varying gain and loss¡ªis tuned.

\begin{figure}
	\centering
	\includegraphics[width=1\columnwidth]{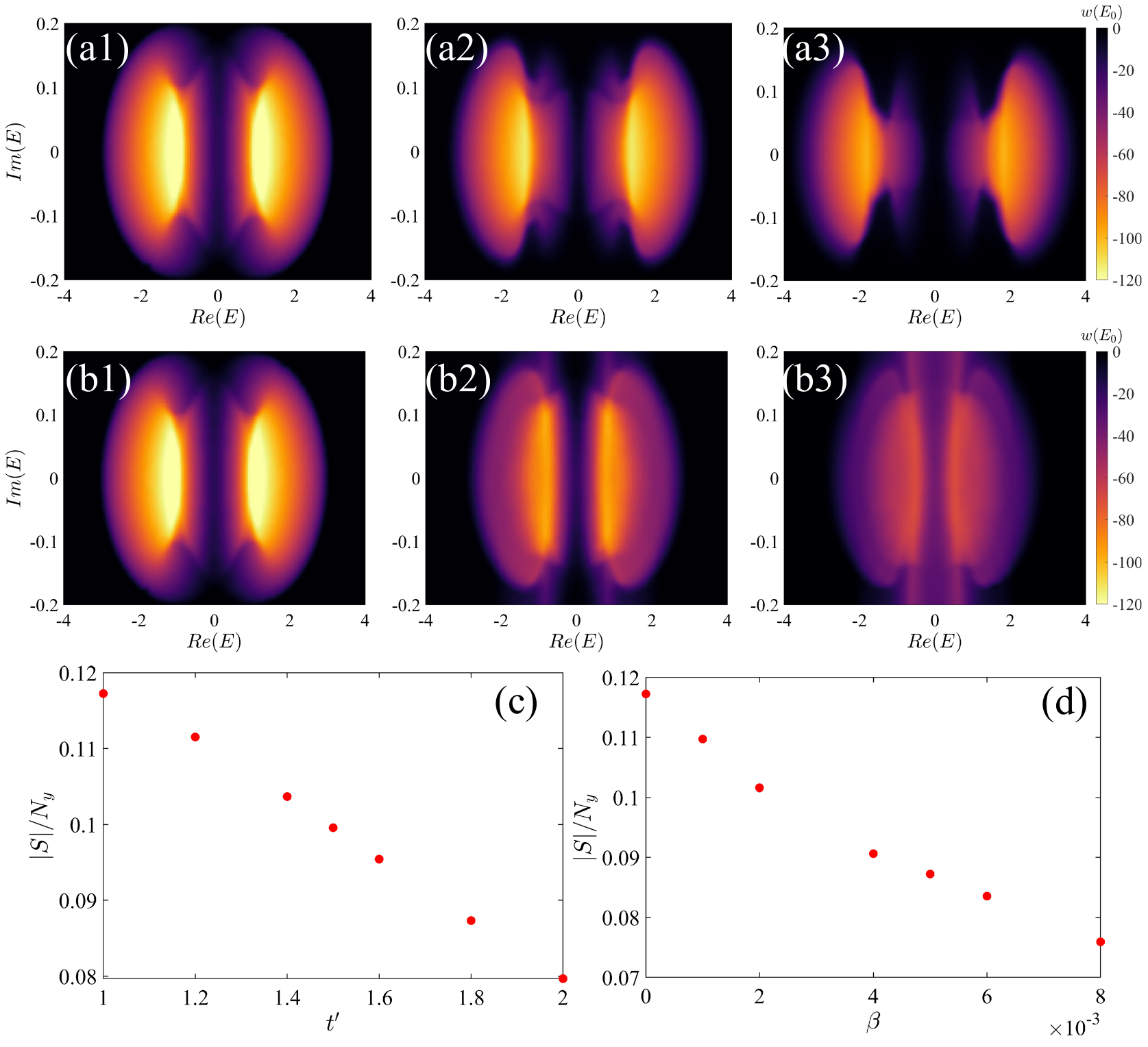}
    \caption{
     (a1)¨C(a3) Distribution of the spectral winding number in the complex energy plane (color scale) for a honeycomb lattice with MIS at $t'=1$, $1.5$, and $2$, respectively.
     (b1)¨C(b3) Corresponding results for a honeycomb lattice with spatially varying gain and loss at $\beta=0$, $0.004$, and $0.008$.
     Skin topological area $|S|/N_y$ as a function of $t'$(c) and $\beta$(d).
     Lattice size $N_y=300$, with $\delta_x=0.1$ and $\delta_y=0$.
     }
	\label{fig3}
\end{figure}

Figs~\ref{fig3}(a1¨Ca3) show the distribution of the spectral winding number in the complex energy plane for increasing PMF strength induced by MIS.
In the absence of the PMF ($t'=1$), a sizable region of the complex plane exhibits nonzero winding numbers, indicating NHSE along the $x$ direction.
As $t'$ increases, the regions with large winding numbers gradually shrink and shift away from zero energy.
For sufficiently strong PMFs, the winding number in the vicinity of zero energy becomes vanishingly small,
and the overall magnitude of the winding number is strongly reduced throughout the complex plane. Importantly, within the point-gap framework of non-Hermitian band theory,
the existence of skin modes along the $x$ direction is associated with the presence of reference energies $E_0$ that yield nonzero winding numbers.
Therefore, the continuous reduction and eventual disappearance of nonzero-winding regions in the complex energy plane
provide direct spectral evidence that the PMF suppresses the NHSE along the $x$ direction.

To quantify this suppression globally, we employ the skin topological area $S$ introduced in Ref.~\cite{LuMing_2021_PRL_Magnetic_Suppression},
which is defined as a weighted integral of the winding number over the complex energy plane.
As shown in Fig.~\ref{fig3}(c), the normalized skin topological area $|S|/N_y$ decreases monotonically with increasing $t'$,
demonstrating a gradual weakening of the NHSE under stronger PMFs.

A similar suppression mechanism is observed for the PMF generated by spatially varying gain and loss.
Figures~\ref{fig3}(b1)¨C(b3) display the winding-number distributions for increasing $\beta$.
Although the imaginary parts of the eigenenergies become larger with increasing $\beta$ due to enhanced gain and loss,
the regions supporting nonzero winding numbers are consistently reduced.
This trend is quantitatively confirmed by the monotonic decrease of $|S|/N_y$ as a function of $\beta$, shown in Fig.~\ref{fig3}(d).

Taken together, these results demonstrate that PMFs¡ªrealized either through MIS or through spatially modulated gain and loss¡ªeffectively suppress the NHSE along the $x$ direction in honeycomb lattices.

\subsection{Pseudomagnetic suppression of the NHSE along the $y$ direction}
\label{Sec3_3}

We now turn to the suppression of the NHSE when the skin accumulation occurs along the $y$ direction.
In contrast to the $x$-direction case discussed in Sec.~\ref{Sec3_2},
the spectral winding number and the associated skin topological area are no longer applicable here.
This is because the presence of a PMF explicitly breaks translational symmetry along $y$,
rendering momentum $k_y$ ill-defined and precluding a point-gap topology analysis in that direction. Instead, the NHSE along $y$ must be diagnosed directly from the spatial distribution of eigenstates under OBC.
To this end, we impose PBC along $x$ and OBC along $y$,
and characterize the localization properties of eigenstates using both numerical and analytical approaches.
\begin{figure}
	\centering
	\includegraphics[width=1\columnwidth]{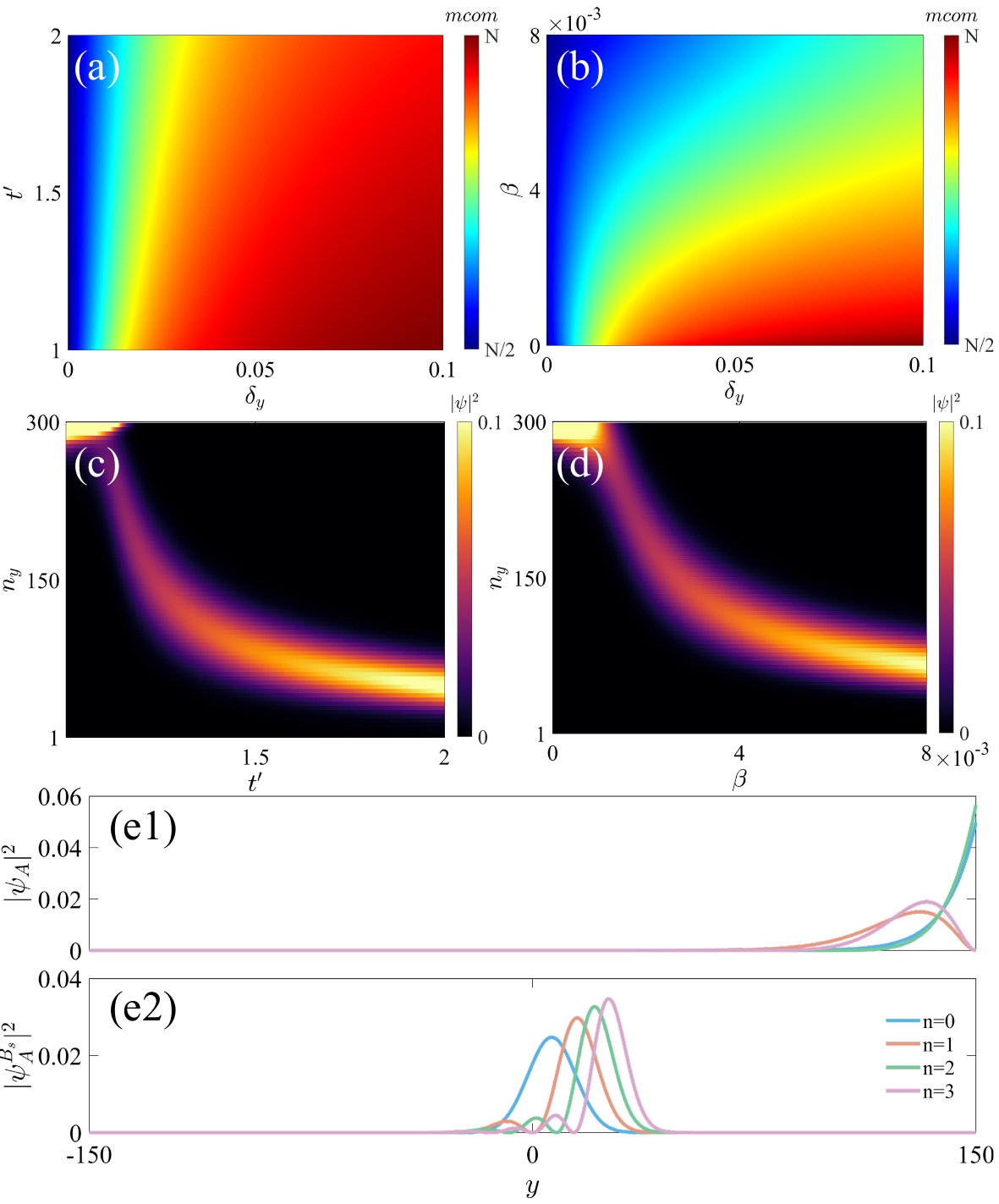}
    \caption{
     \textit{mcom} in the $(\delta_y , t')$ plane (a) and $(\delta_y , \beta)$ plane (b).
     Spatial distributions of an eigenstate belonging to the first pseudo-Landau level as a function of $t'$ (c) and $\beta$ (d),
     illustrating the transition from a boundary-localized skin mode to a bulk-localized mode.
     (e1) Amplitude distribution $|\psi^n_A|^2$ ($n=0,1,2,3$ denote the lowest-energy states) in the absence of a PMF, showing skin localization near the boundary.
     (e2) Corresponding analytical wave functions obtained from Eq.~\eqref{psi_A_y}, demonstrating bulk localization induced by the PMF.
     Parameters: $N_y=300$, $q_x=0$, $\tilde{\delta}=0.05$, $B_s=0.008$.
     }
	\label{fig4}
\end{figure}

To quantitatively diagnose the NHSE along $y$, we introduce the mean center of mass (\textit{mcom}) of the eigenstate amplitudes~\cite{JinWeiWu_2025_PRL_Anderson_Delocalization,Molignini_2023_PRRes_disordered},
which captures the average spatial location of eigenstates along the open direction.
It is defined as
\begin{equation}\label{mcom1_re}
\begin{split}
\mathrm{mcom}=
\frac{\sum_{j=1}^{N_y} j\,\langle\mathcal{A}(j)\rangle}
{\sum_{j=1}^{N_y}\langle\mathcal{A}(j)\rangle},
\end{split}
\end{equation}
where
\begin{equation}\label{mcom2_re}
\begin{split}
\langle\mathcal{A}(j)\rangle=
\left\langle
\frac{1}{N}\sum_{n=1}^{N}
\sum_{\sigma=1}^{4}
\left|\psi_{n}^{(\sigma)}(j)\right|^{2}
\right\rangle .
\end{split}
\end{equation}
Here, $\sigma$ labels the four sublattices, and $\langle\cdot\rangle$ denotes averaging over $k_x$.
For a system with $N_y$ sites along $y$, values of $\mathrm{mcom}$ close to $1$ or $N_y$ indicate boundary localization characteristic of the NHSE,
whereas $\textit{mcom}\approx N_y/2$ signals bulk-localized states.

Figs~\ref{fig4}(a,b) show the calculated \textit{mcom} for PMFs induced by MIS and by spatially varying gain and loss, respectively.
For a fixed nonreciprocal hopping strength $\delta_y$,
increasing either $t'$ or $\beta$ systematically drives \textit{mcom} from boundary values toward the bulk center.
This behavior demonstrates that both types of PMFs effectively counteract the NHSE accumulation along $y$.
To visualize this suppression mechanism more explicitly,
we track the spatial profiles of representative eigenstates near the valley centers at
$k_x = 4\pi/(3a)$ [Fig.~\ref{fig4}(c)] and $k_x = 2\pi/(3a)$ [Fig.~\ref{fig4}(d)].
The selected states correspond, in the Hermitian limit, to the first pseudo-Landau level.
In the absence of a PMF ($t'=1$ or $\beta=0$),
these states exhibit exponential localization at the boundary, characteristic of $y$-direction skin modes.
As the PMF strength increases,
the localization gradually weakens and the wave-function maximum shifts into the bulk,
eventually forming a bulk-confined Landau-type mode.

The suppression effect can also be investigated analytically within the long-wavelength limit.
We consider the case of a PMF generated by MIS, starting from Eq.~\eqref{G_h_q} and introducing a non-Hermitian term $\delta_y$,
we examine the system's low-energy behavior.
In the absence of a PMF, expanding the Hamiltonian near the $\mathbf{K}$ valley yields the low-energy effective form
$h(\mathbf{q},\delta_y)=\hbar v_F[q_x\sigma_x+(q_y+i\tilde{\delta})\sigma_y]$,
where $\tilde{\delta}=2\delta_y/\sqrt{3}at$.
Taking the trial wave function $\psi=(\psi_A,\psi_B)^{\mathrm{T}}$ and imposing OBC $\psi(y=0)=\psi(y=N_y)=0$,
the $A$-sublattice component of the wave function is obtained as
$\psi_A(y)\propto e^{\tilde{\delta} y}\sin\!\left( \frac{n\pi y}{L} \right)$ [cf. Appendix~\ref{AppB}].

To incorporate the effect of the PMF, we perform the Peierls substitution
$\hbar \mathbf{q}\rightarrow \mathbf{p}-q\mathbf{A}_p$,
where $\mathbf{p}$ is the canonical momentum.
Under this substitution, considering the pseudogauge field $\mathbf{A}_s=(B_s y,0)$ and taking $q=1$,
the effective Hamiltonian becomes
\begin{equation}\label{hq_Bs_dy0}
\begin{split}
h_{\text{eff}}=
\begin{pmatrix}
    0 & q_x-B_s y-\frac{\partial}{\partial y}+\tilde{\delta} \\
    q_x-B_s y+\frac{\partial}{\partial y}-\tilde{\delta} & 0 \\
\end{pmatrix},
\end{split}
\end{equation}
where the $y$-OBC has been incorporated and the substitution $q_y=-i\hbar\,\partial/\partial y$ has been applied.
From Eq.~\eqref{hq_Bs_dy0}, the wave function $\psi^{B_s}_A(y)$ for $B_s\neq 0$ under OBC can be expressed as
\begin{equation}\label{psi_A_y}
\begin{split}
\psi^{B_s}_{A,n}(y)\propto e^{\tilde{\delta} y}
e^{-\tfrac{(y-y_0)^2}{2l_{B_s}^2}}
\mathcal{H}_n\!\left(\frac{y-y_0}{l_{B_s}}\right),
\end{split}
\end{equation}
where $y_0=y-q_x/B_s$, $l_{B_s}=B_s^{-1/2}$, and $\mathcal{H}_n$ is the $n$th Hermite polynomial [cf. Appendix~\ref{app_B}]. The spatial distributions of the four lowest-energy states are shown in Fig.~\ref{fig4}(e1) for $B_s=0$ and Fig.~\ref{fig4}(e2) for $B_s\neq 0$, with $y\in[-150,150]$.
When $B_s=0$, the non-Hermitian term $\delta_y$ induces localization of the wave function near the boundary ($y\!\to\!150$).
In contrast, for $B_s\neq 0$, the Hermite polynomial components induced by the PMF in Eq.~\eqref{psi_A_y} redistribute the skin modes from the boundary into the bulk ($y\!\to\!0$).
Therefore, within the low-energy limit, we conclude that the PMF suppresses the NHSE along the $y$-direction.

\begin{figure}
	\centering
	\includegraphics[width=1\columnwidth]{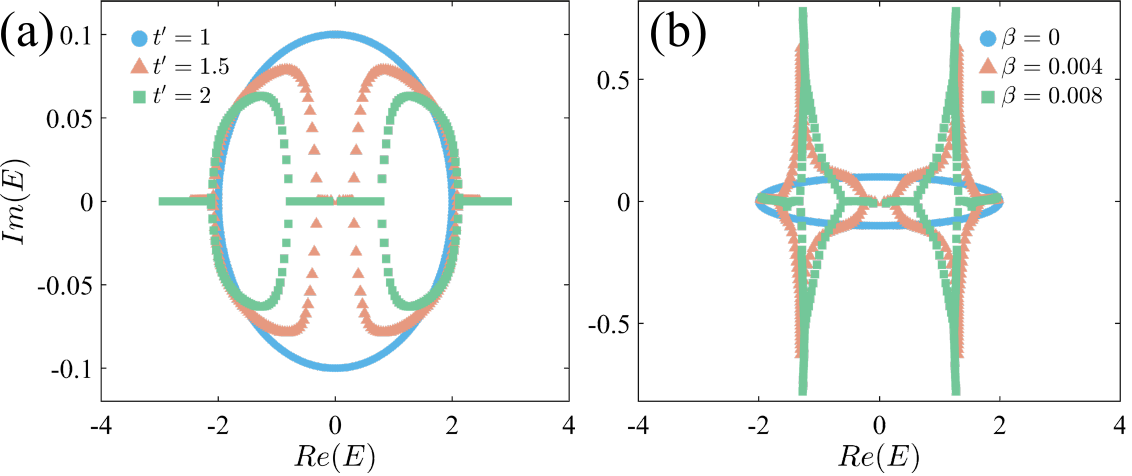}
    \caption{
    Complex energy spectra under $y$-PBC for (a) MIS-induced and (b) gain--loss-induced PMFs at different field strengths (color coded).
    Parameters: $N_y=300$, $k_x=2\pi/(3a)$, $\delta_y=0.1$.
    }
	\label{fig5}
\end{figure}

Finally, we examine the complex energy spectra under $y$-PBC to complement the real-space analysis.
As shown in Fig.~\ref{fig5}(a),
for the MIS-induced PMF, the spectral area in the complex plane contracts with increasing field strength,
and the band-edge eigenvalues approach the real axis,
consistent with the behavior of Landau levels under a real magnetic field~\cite{Shaokai_2022_PRB_Cyclotron}.
For the gain¨Closs¨Cinduced PMF [Fig.~\ref{fig5}(b)],
although gain and loss initially broaden the spectrum along the imaginary axis,
the spectral area near zero energy is progressively reduced as the PMF increases.
These spectral trends further corroborate the real-space signatures of NHSE suppression.

In summary, the \textit{mcom} analysis, direct wave-function profiles, analytical low-energy solutions,
and spectral evolution under $y$-PBC consistently demonstrate that PMFs effectively suppress the NHSE along the $y$ direction.

\section{Suppression of the NHSE by PMFs with different edge terminations}
\label{Sec4}
In the previous sections, we have demonstrated that PMFs,
generated either by MIS or by spatially varying gain and loss,
can effectively suppress the NHSE in honeycomb nanoribbons with zigzag edges.
However, the honeycomb lattice supports several inequivalent boundary terminations,
including bearded, armchair, and twig edges,
which are known to host distinct edge states and boundary spectra in Hermitian systems
\cite{CastroNeto_2009_RMP,Nakada_1996_PRB_EdgeState,Kohmoto_2007_PRB_Zeromodes,XiaShiqi_2023_PRL_Photonic,GaoFeng_2024_PRAppl_Visualization}.
It is therefore essential to examine whether the PMF-induced suppression of the NHSE
is robust against changes in boundary geometry. In this section, we extend our analysis to honeycomb nanoribbons with bearded, armchair, and twig edges.
We show that, despite quantitative differences in band structures and edge-state dispersions,
the qualitative mechanism by which PMFs suppress the NHSE remains unchanged,
demonstrating the generality of our conclusions.
\\

\subsection{Bearded edges}
\begin{figure}
	\centering
	\includegraphics[width=1\columnwidth]{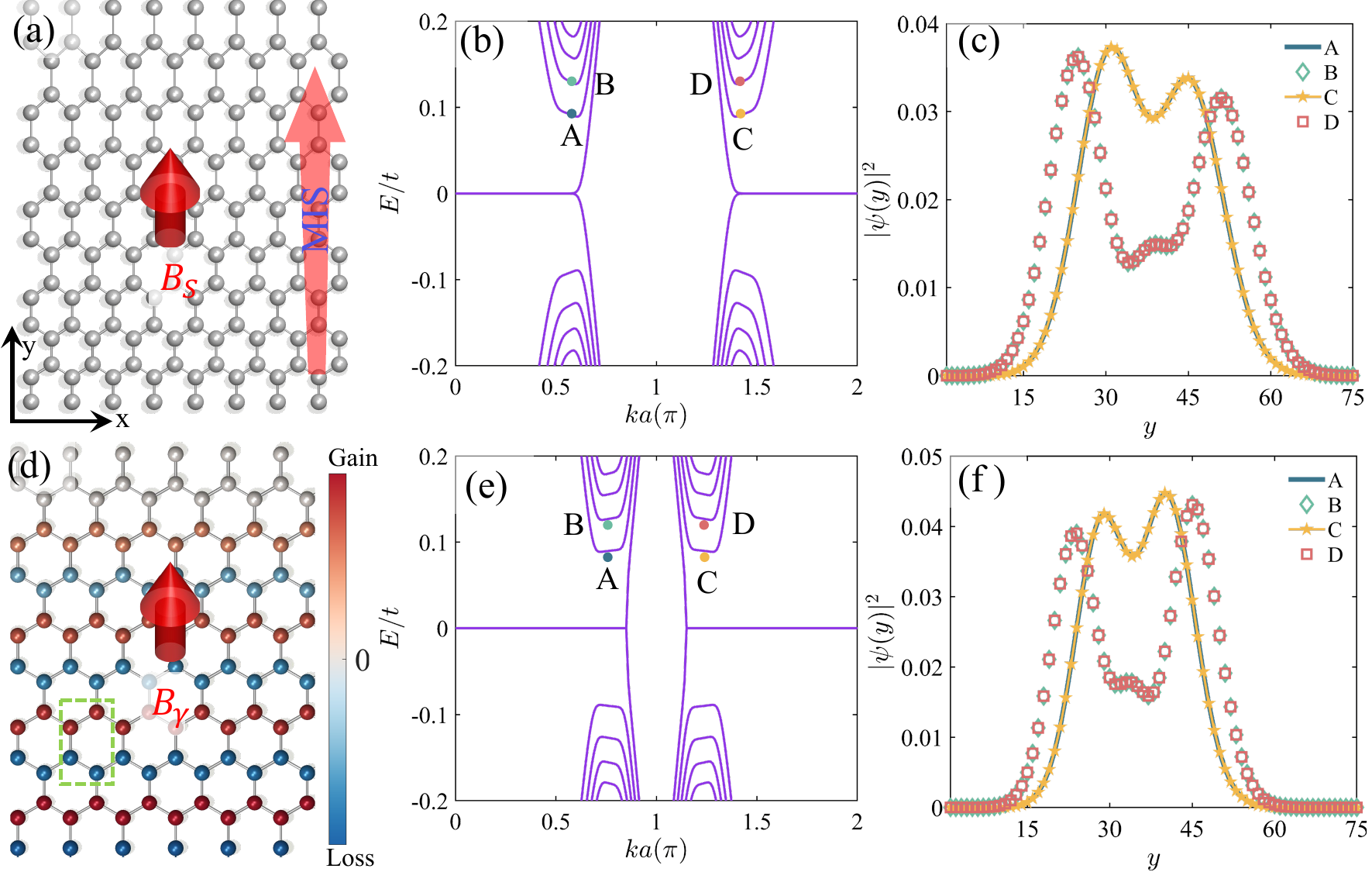}
    \caption{
    Realization of a uniform PMF $B_s$ in a honeycomb lattice with bearded edges via MIS (a),
    and generation of a PMF via spatially varying gain and loss strength $\gamma$ along the $y$ direction (d).
    Nonreciprocal hopping is defined in the same manner as in Fig.~\ref{fig1}(a,b).
    (b,e) Energy bands corresponding to (a,d) for zero nonreciprocal hopping ($\delta_x=\delta_y=0$).
    (c,f) Wave-function distributions $|\psi(y)|^2$ calculated at the marked points in (b,e).
    Parameters: $N_y=298$, $t=1$, $\delta_x=\delta_y=0$, $t'=1.5$ in (b,c), and $\beta=0.005$ in (e,f).
    }
	\label{fig6}
\end{figure}

A bearded edge can be viewed as a zigzag edge decorated with additional dangling sites.
Figs~\ref{fig6}(a,d) show honeycomb nanoribbons terminated by two bearded edges.
Following the zigzag-edge case, we first introduce a MIS along the $y$ direction.
The resulting band structure, shown in Fig.~\ref{fig6}(b), clearly exhibits dispersive pseudo-Landau levels.
The corresponding eigenstates are localized in the bulk,
as confirmed by the spatial profiles $|\psi(y)|^2$ in Fig.~\ref{fig6}(c).

Compared to the zigzag geometry, a notable difference is that the zero-energy mode
now appears near $ka=0$ rather than around $ka=\pi$.
This shift reflects the modified boundary termination but does not affect
the bulk localization induced by the PMF.
We next consider PMFs generated by spatially varying gain and loss [Fig.~\ref{fig6}(d)].
As shown in Fig.~\ref{fig6}(e), pseudo-Landau levels persist in this non-Hermitian realization,
and the associated eigenstates remain bulk localized [Fig.~\ref{fig6}(f)].
These results demonstrate that both PMF-generation schemes remain effective
for honeycomb ribbons with bearded edges.

\begin{figure*}
	\centering
	\includegraphics[width=2\columnwidth]{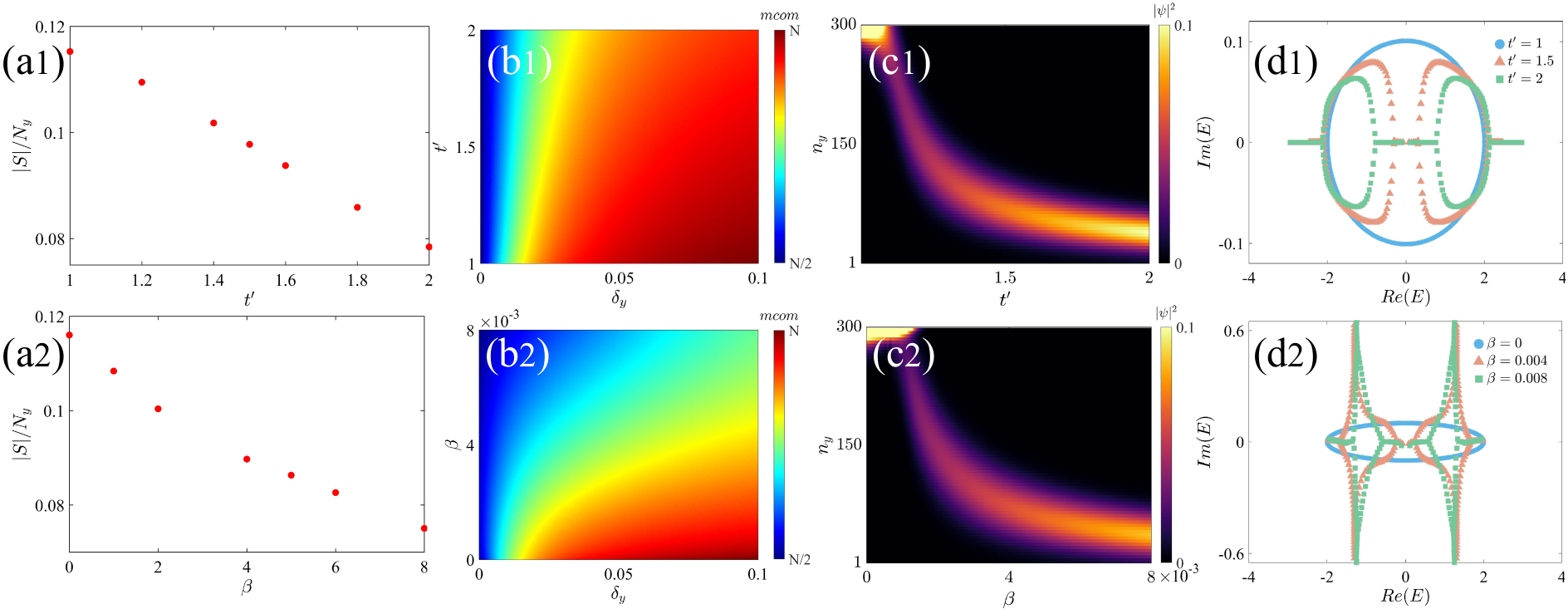}
    \caption{
    Suppression of the NHSE in honeycomb ribbons with bearded edges.
    (a1,a2) Normalized skin topological area $|S|/N_y$ as a function of $t'$ and $\beta$,
    evaluated within $\text{Re}(E)\in[-4,4]$ and $\text{Im}(E)\in(-0.2,0.2)$.
    (b1,b2) \textit{mcom} in the $(\delta_y,t')$ and $(\delta_y,\beta)$ planes.
    (c1,c2) Spatial distributions of the first pseudo-Landau-level eigenstate as a function of $t'$ and $\beta$.
    (d1,d2) Complex energy spectra under $y$-PBC for MIS-induced and gain--loss-induced PMFs.
    Parameters: $\delta_x=0.1$ in (a1,a2), $\delta_y=0.1$ in (b1--d2),
    $k_x=2\pi/(3a)$ in (d1,d2), and $N_y=298$.
    }
	\label{fig7}
\end{figure*}

To quantify the suppression of the NHSE, we first consider nonreciprocal hopping along the $x$ direction
($\delta_x\neq0$, $\delta_y=0$).
The spectral winding number defined in Eq.~\eqref{Winding_num} is evaluated in the complex-energy plane,
from which the normalized skin topological area is extracted.
As shown in Figs.~\ref{fig7}(a1) and~\ref{fig7}(a2),
the skin topological area decreases monotonically with increasing PMF strength
for both MIS-induced and gain--loss-induced PMFs,
signaling a progressive suppression of the NHSE along the $x$ direction.

For skin modes along the $y$ direction ($\delta_y\neq0$),
we employ the \textit{mcom} under $y$-OBC.
The phase diagrams shown in Figs.~\ref{fig7}(b1,b2)
demonstrate that increasing the PMF strength drives the eigenstates
from boundary-localized skin modes toward bulk-localized states.
This behavior is further corroborated by the real-space profiles of the 1-st pseudo-Landau-level eigenstates
in Figs.~\ref{fig7}(c1,c2).
Consistently, under $y$-PBC the area enclosed by the complex-energy spectrum
shrinks with increasing PMF strength
[Figs.~\ref{fig7}(d1,d2)]. These results establish that PMFs also suppress the NHSE
in honeycomb nanoribbons with bearded edges,
in a manner that is directly analogous to the case of zigzag edges.

\subsection{Armchair edges}

\begin{figure}
	\centering
	\includegraphics[width=1\columnwidth]{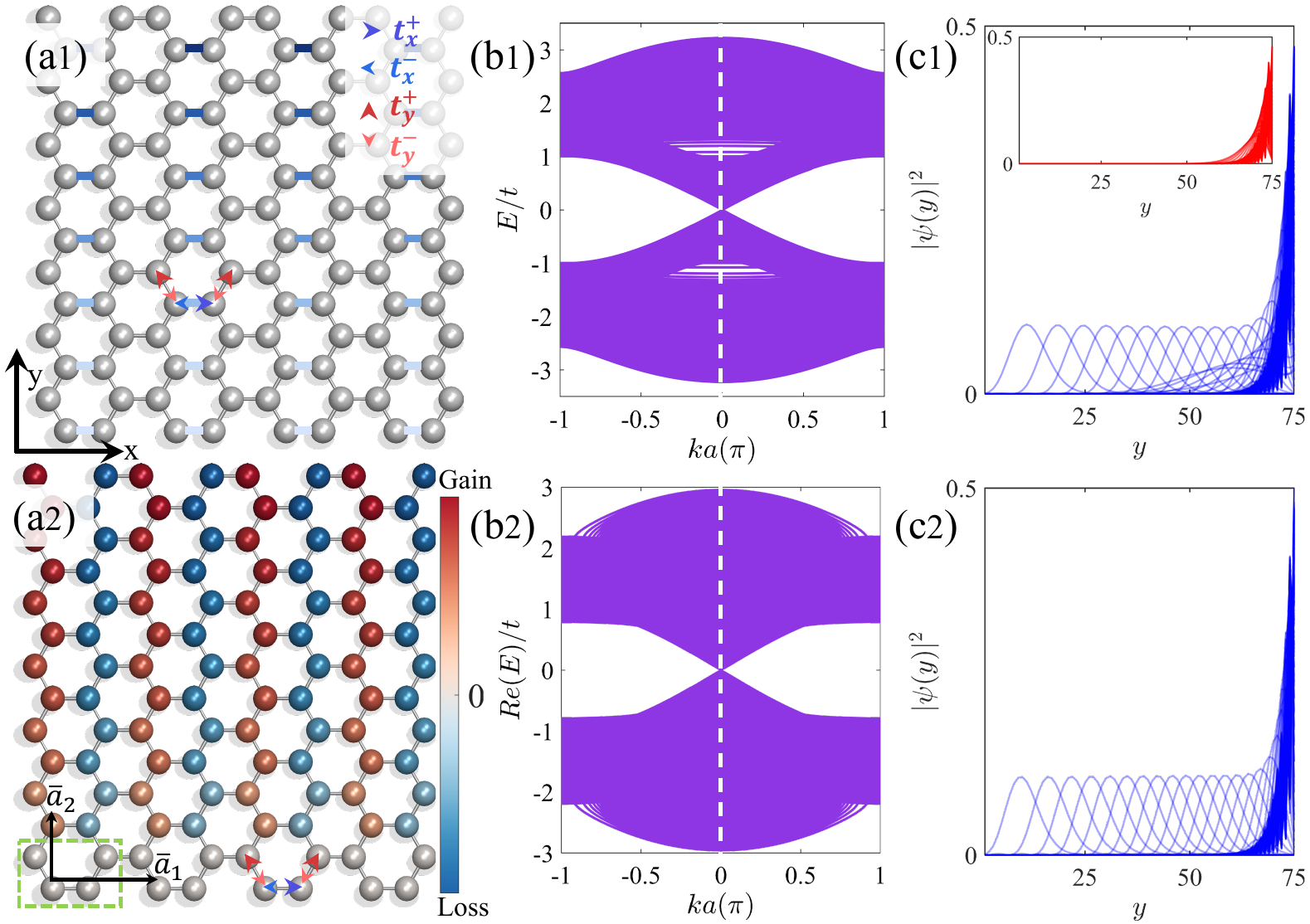}
    \caption{
    Non-Hermitian honeycomb nanoribbons with armchair edges.
    (a1) Spatially uniform hopping modulation within each unit cell,
    varying linearly along the $y$ direction, generating a PMF via MIS.
    (a2) Spatially varying gain and loss along the $y$ direction indicated by the color gradient,
    generating a PMF in the non-Hermitian setting.
    Nonreciprocal hopping $\delta_y$ ($\delta_x$) is indicated by (light) red and (light) purple arrows,
    superimposed on the $y$ ($x$) direction hopping.
    Black arrows denote the primitive lattice vectors $\bar{\mathbf{a}}_1$ and $\bar{\mathbf{a}}_2$,
    and the green dashed box marks the unit cell.
    (b1,b2) Energy bands corresponding to (a1,a2) with $\delta_x=0$ and $\delta_y=0.1$.
    (c1,c2) Effective wave-function profiles $|\psi(y)|^2$ at $k=0$
    [vertical dashed lines in (b1,b2)].
    The inset in (c1) shows the non-Hermitian case without PMF for comparison.
    Parameters: $N_y=300$, $t=1$, $\delta_x=0$, $\delta_y=0.1$,
    $t'=1.5$ in (b1,c1), and $\beta=0.005$ in (b2,c2).
    }
	\label{fig8}
\end{figure}

We next consider non-Hermitian honeycomb nanoribbons with armchair edges,
as illustrated in Figs.~\ref{fig8}(a1,a2).
In this geometry, translational symmetry along the $x$ direction is preserved,
and the $K$ and $K'$ valleys are folded onto each other in the one-dimensional Brillouin zone.
As a result, pseudo-Landau levels cannot be directly identified from the band alone. To characterize the interplay between the PMF and the NHSE,
we therefore analyze the real-space properties of the eigenstates at fixed momentum.
Specifically, we focus on $k=0$, marked by the vertical dashed lines in Figs.~\ref{fig8}(b1,b2).
By summing the wave-function intensities over the four lattice sites within each unit cell,
we obtain an effective spatial profile along the $y$ direction. As shown in Fig.~\ref{fig8}(c1), in the absence of a PMF,
nonreciprocal hopping ($\delta_y\neq0$) leads to strong boundary localization,
a hallmark of the NHSE.
Upon introducing a PMF via MIS, this boundary accumulation is suppressed,
and some eigenstates become redistributed into the bulk.
An analogous suppression of the NHSE is observed when the PMF is generated
by spatially varying gain and loss, as shown in Fig.~\ref{fig8}(c2).

\begin{figure}
	\centering
	\includegraphics[width=1\columnwidth]{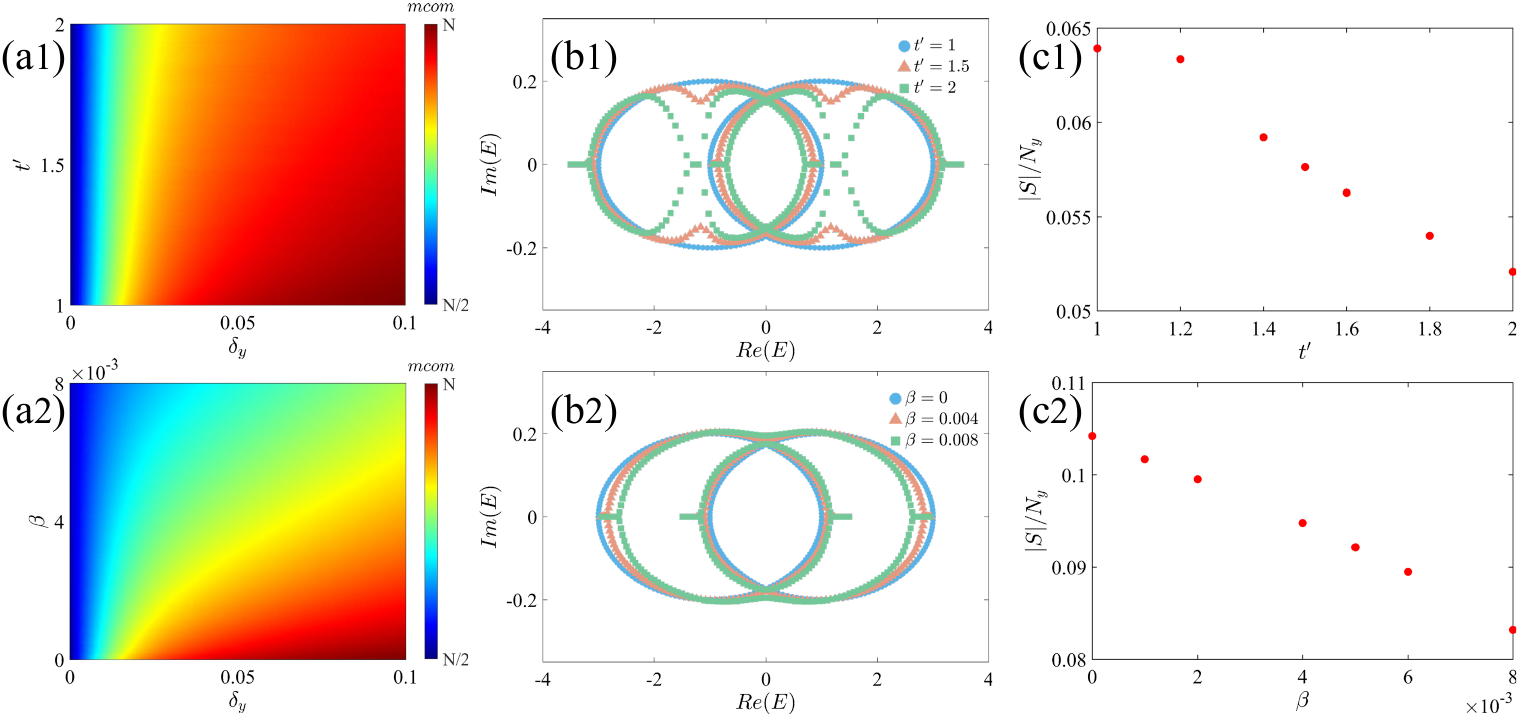}
    \caption{
    Suppression of the NHSE in armchair-edged ribbons.
    (a1,a2) \textit{mcom} under $y$-OBC
    in the $(\delta_y,t')$ and $(\delta_y,\beta)$ planes.
    (b1,b2) Complex energy spectra under $y$-PBC
    for MIS-induced and gain--loss-induced PMFs at different strengths (color coded).
    (c1,c2) Skin topological area $ |S|/N_y$ as a function of $t'$ with $\text{Re}(E)\in [-4,4]$ and $\text{Im}(E)\in (-0.2,0.2)$ (c1) and $\beta$ with $\text{Re}(E)\in [-1,1]$ and $\text{Im}(E)\in (-0.01,0.01)$(c2).  Calculation parameters: $\delta_y=0.1$ in (a1-b2), $\delta_x=0.1$ in (c1-c2), $k_x=0$ in (b1,b2) and $ N_y = 300$.
    }
	\label{fig9}
\end{figure}

To further quantify this behavior,
we compute the \textit{mcom} under $y$-OBC,
shown in Figs.~\ref{fig9}(a1,a2).
For fixed nonreciprocity $\delta_y$, increasing the PMF strength
drives the \textit{mcom} from values close to the system boundaries toward $N_y/2$,
indicating a progressive delocalization of skin modes into the bulk. Consistently, under $y$-PBC the complex-energy spectra
shrink
as the PMF strength increases
[Figs.~\ref{fig9}(b1,b2)].
For NHSE along the $x$ direction,
the normalized skin topological area extracted from the spectral winding number
also decreases monotonically with increasing PMF strength
[Figs.~\ref{fig9}(c1-c2)]. These results demonstrate that
PMFs robustly suppress the NHSE in honeycomb nanoribbons with armchair edges.

\subsection{Twig edges}

\begin{figure}
	\centering
	\includegraphics[width=1\columnwidth]{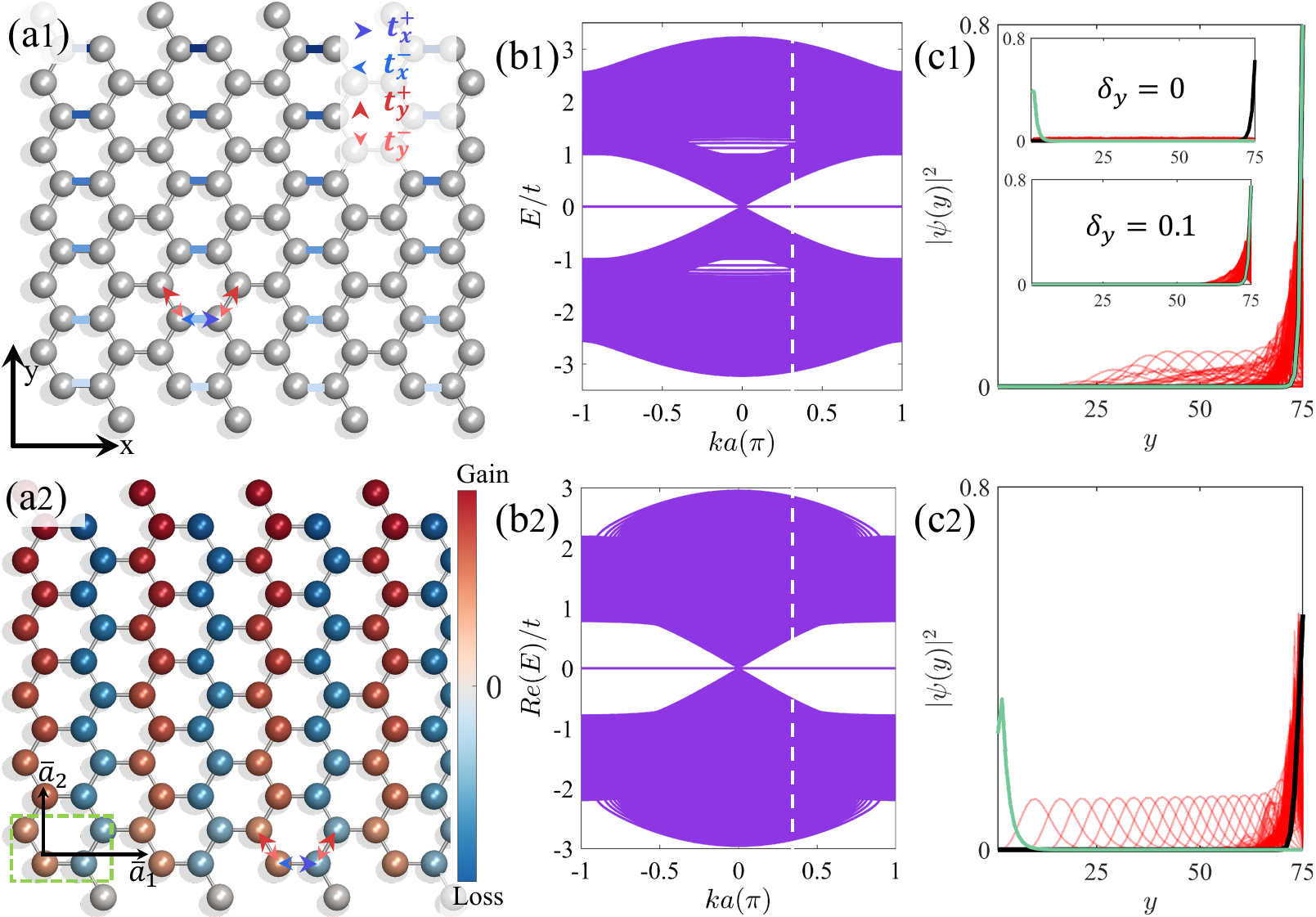}
    \caption{
    Non-Hermitian honeycomb nanoribbons with twig edges.
    (a1) Spatially uniform hopping modulation along the $y$ direction (MIS-induced PMF).
    (a2) Spatially varying gain and loss along the $y$ direction (gain--loss-induced PMF).
    Nonreciprocal hopping $\delta_x$ ($\delta_y$) is indicated by (light) red and (light) purple arrows
    superimposed on the $x$ ($y$) direction hopping.
    Black arrows denote the primitive vectors $\bar{\mathbf{a}}_1$ and $\bar{\mathbf{a}}_2$,
    and the green dashed box marks the unit cell.
    (b1,b2) Energy bands corresponding to (a1,a2) with $\delta_x=0$ and $\delta_y=0.1$.
    (c1,c2) Effective wave-function profiles $|\psi(y)|^2$ at $ka=0.3\pi$
    [vertical dashed lines in (b1,b2)].
    Insets in (c1) show the Hermitian and non-Hermitian cases without PMF.
    Parameters: $N_y=296$, $t=1$, $\delta_x=0$, $\delta_y=0.1$,
    $t'=1.5$ in (b1,c1), and $\beta=0.005$ in (b2,c2).
    }
	\label{fig10}
\end{figure}

We finally consider honeycomb nanoribbons with twig edges,
a special boundary geometry obtained by attaching dangling sites
to an armchair edge at the two longitudinal ends of the ribbon
[Figs.~\ref{fig10}(a1,a2)].
This edge termination is known to host flat bands originating
from the dangling sites~\cite{XiaShiqi_2023_PRL_Photonic,GaoFeng_2024_PRAppl_Visualization}.

Following the analysis of armchair-edged ribbons,
we also generate PMFs using two distinct schemes:
a spatially linear hopping modulation along the $y$ direction
[Fig.~\ref{fig10}(a1)]
and spatially varying gain and loss
[Fig.~\ref{fig10}(a2)].
In both cases, translational symmetry along the $x$ direction is preserved,
and the energy spectra under $y$-OBC are shown in
Figs.~\ref{fig10}(b1) and~\ref{fig10}(b2).
In contrast to the armchair-edge geometry,
flat bands are clearly visible in the twig-edge ribbons.

To analyze the real-space properties,
we examine eigenstates at $ka=0.3\pi$,
marked by the vertical dashed lines in the band structures.
The wave-function profiles $|\psi(y)|^2$
are obtained by summing the amplitudes over the four sites within each unit cell,
while treating the boundary dangling sites separately.
As shown in Fig.~\ref{fig10}(c1),
flat-band states are predominantly localized near the boundaries,
reflecting their geometric origin.
Introducing nonreciprocal hopping alone alters
this edge localization, as shown in the inset of Fig.~\ref{fig10}(c1).
It is worth noting that the spatial distribution of these edge states depends on the crystal momentum,
here we present a representative case at $ka = 0.3\pi$.
Upon further introducing the PMF,
the localization characteristics of the two degenerate flat-band edge states exhibit distinct manifestations depending on the induction mechanism.
For the MIS-induced PMF, the edge states are predominantly influenced by non-Hermitian effects, remaining localized at a single boundary of the system [Fig.~\ref{fig10}(c1)].
In contrast, under the PMF generated via gain-loss modulation, the two edge states recover a spatial distribution consistent with the Hermitian limit, with one state localized at the upper boundary and the other at the lower boundary [Fig.~\ref{fig10}(c2)].

\begin{figure}
	\centering
	\includegraphics[width=1\columnwidth]{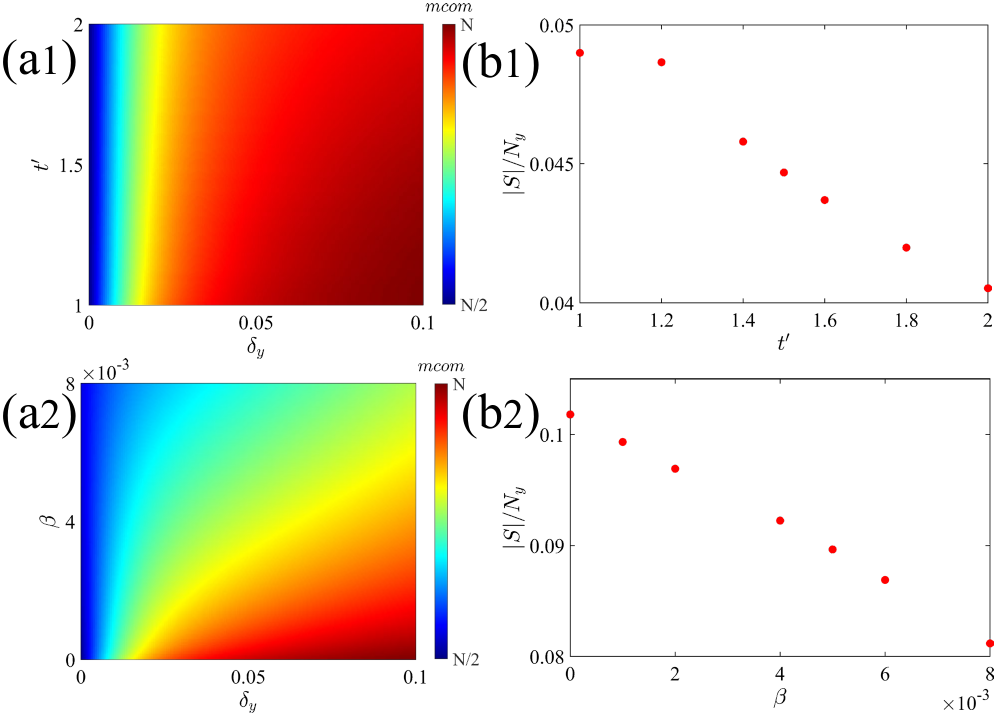}
    \caption{
    Suppression of the NHSE in twig-edged ribbons.
    (a1,a2) \textit{mcom} under $y$-OBC
    in the $(\delta_y,t')$ and $(\delta_y,\beta)$ planes.
    Skin topological area $ |S|/N_y$ as a function of $t'$ with $\text{Re}(E)\in [-4,4]$ and $\text{Im}(E)\in (-0.2,0.2)$ (b1) and $\beta$ with $\text{Re}(E)\in [-1,1]$ and $\text{Im}(E)\in (-0.01,0.01)$(b2).
    Parameters: $\delta_y=0.1$ in (a1,a2), $\delta_x=0.1$ in (b1,b2),
    and $N_y=296$.
    }
	\label{fig11}
\end{figure}
Crucially, for bulk states that exhibit boundary localization due to the NHSE,
the PMF effectively suppresses the skin accumulation
and redistributes these states into the bulk,
consistent with the behavior observed for other edge terminations.
To quantitatively characterize the suppression of the NHSE,
we compute the \textit{mcom} under $y$-OBC,
shown in Figs.~\ref{fig11}(a1,a2).
For fixed nonreciprocity $\delta_y$,
increasing the PMF strength drives the \textit{mcom}
from values close to the system boundaries toward $N_y/2$,
indicating a progressive delocalization of skin modes into the bulk.
This behavior is observed for both PMF-generation mechanisms.
For NHSE along the $x$ direction,
we further evaluate the spectral winding number
and extract the normalized skin topological area.
As shown in Figs.~\ref{fig11}(b1,b2),
the topological area decreases monotonically with increasing PMF strength,
demonstrating that the PMF also suppresses the NHSE in this direction.

Overall, despite the presence of flat-band edge states intrinsic to the twig-edge geometry,
PMFs generated either by hopping gradients or by spatially varying gain and loss
robustly suppress the NHSE in honeycomb nanoribbons with twig edges.
Together with the results for zigzag, bearded, and armchair edges,
these findings establish the generality of the PMF-induced suppression mechanism.

\section{Conclusion and Outlook}
\label{Sec5}

In conclusion, we have demonstrated that PMFs,
generated either by MIS or by spatially varying gain and loss,
provide an efficient and universal mechanism for suppressing the NHSE
in honeycomb lattices.
To establish the robustness of this mechanism,
we systematically investigated honeycomb nanoribbons with representative boundary terminations,
including zigzag, bearded, armchair, and twig edges.
Despite their distinct edge geometries and spectral characteristics,
all cases exhibit qualitatively consistent behavior,
demonstrating that PMF-induced suppression of the NHSE is a generic phenomenon
that does not rely on boundary terminations.

For NHSE along the $x$ direction,
we characterized the skin effect using the spectral winding number
and the associated skin topological area in the complex energy plane.
Both quantities decrease monotonically with increasing PMF strength
for PMFs induced by either scheme,
indicating a progressive weakening and eventual suppression of the skin accumulation.
For NHSE along the $y$ direction under OBC,
we employed the \textit{mcom} together with real-space wave-function analysis.
The results show that PMFs continuously drive boundary-localized skin modes back into the bulk,
a behavior that is further captured by an effective low-energy continuum description.
Under PBC along the $y$ direction,
the complex-energy spectra exhibit a systematic contraction with increasing PMF strength,
accompanied by complex-to-real transitions of eigenvalues near the band edges.
This spectral evolution closely parallels that induced by a real magnetic field~\cite{Shaokai_2022_PRB_Cyclotron},
providing complementary confirmation of PMF-induced NHSE suppression.

Compared with other strategies for controlling the NHSE,
such as real magnetic fields~\cite{Shaokai_2022_PRB_Cyclotron,LuMing_2021_PRL_Magnetic_Suppression},
graded potentials~\cite{ChaoXu_2025_PRB_Controllable},
or electric fields~\cite{YiPeng_2022_PRB_Manipulating},
PMFs offer several distinctive advantages.
In particular, they can reach substantially larger effective field strengths~\cite{guinea_NP_2010_Strain,levy_Science_2010_strain}
and can be implemented with high flexibility in artificial platforms.
In acoustic systems, NHSE and PMFs can be realized via engineered nonreciprocal couplings
and spatially modulated structures~\cite{zhangXiujuan_NC_2021observation,ZhangLi_2021_NC_acoustic,GaoHe_2022_PRB_AnomalousFloquet,GuZhongming_2022_NC_Transient,YangZhaoju_2017_PRL_Strain_Induced,Abbaszadeh_2017_PRL_Sonic,WenXinhua_2019_NP_Acoustic_Landau,YanMouDengweiying_2021_PRL_Pseudomagnetic}.
In photonic platforms, they can be implemented through complex
and spatially varying refractive indices~\cite{ZhuXueyi_PRRes_2020_Photonic,LinZekun_2021_OE_Square_root,Yokomizo_2022_PRRes_photoniccrystals,rechtsman_Np_2013_strain,ZhangXiao_2025_PRL_PhotonicFlatLandau,WangWenhui_2020_PRL_MoireFringe,jamadi_LightSA_2020_Direct,Bellec_2020_Light_SA_ObservationPseudo,Guglielmon_2021_PRA_LandauStrained}.
In electrical circuit realizations~\cite{Imhof_NP_2018_Topolectricalcircuit,Helbig_2020_NP_GeneralizedBBC,LiuShuo_2021_Res_Non_Hermitian,LeeChingHua_2018_Commu_Phy_Topolectrical,teo2024_SciBul_pseudomagnetic},
PMFs can be engineered using negative-impedance converters combined with inductor--capacitor networks.

The theoretical framework and results presented here
provide clear guidance for future experimental explorations
of non-Hermitian phenomena in honeycomb lattices
and can be naturally extended to higher-dimensional non-Hermitian systems.
More broadly, our findings open new possibilities for controlling localization phenomena
and actively manipulating wave dynamics in classical platforms,
including photonic, acoustic, and electrical systems.

\begin{acknowledgments}
We thank Ming Lu for helpful discussions. This work was supported by the National Key Projects for Research and Development of China under Grant No. 2022YFA120470 (W. C.), the National Natural Science Foundation of China under Grant No. 12074172 (W. C.), No. 12222406 (W. C.).
\end{acknowledgments}

\begin{appendix}

\section{Derivation of Pseudo-gauge Potential and Fermi Velocity in Strained Honeycomb Lattice}
\label{AppA}
In this Appendix, we provide the detailed derivation of the strain-induced pseudo-gauge potential $\mathbf{A}$ and the corresponding anisotropic Fermi velocities $v_x$ and $v_y$ in strained honeycomb lattice.
Since the analysis for the $K'$ valley proceeds analogously, we restrict our discussion to the $K$ valley.

Within the nearest-neighbor tight-binding approximation, the Hamiltonian of strained honeycomb lattice in momentum space takes the form
\begin{equation}\label{Hk}
\begin{split}
H_k &=
\begin{pmatrix}
0 & h_{12} \\
h_{12}^* & 0
\end{pmatrix}  \\
&=
\begin{pmatrix}
0 & t_y + t e^{i\mathbf{k}\cdot\mathbf{a}_1} + t e^{i\mathbf{k}\cdot\mathbf{a}_2} \\
t_y + t e^{-i\mathbf{k}\cdot\mathbf{a}_1} + t e^{-i\mathbf{k}\cdot\mathbf{a}_2} & 0
\end{pmatrix},
\end{split}
\end{equation}

The Dirac point in the $K$ valley is located at
\[
\mathbf{K}^{\mathrm{D}}=\left(\frac{4\pi}{3a},0\right),
\]
which satisfies the condition
\begin{equation}\label{h12}
\begin{split}
h_{12}=t+2t\cos\!\left(\frac{a}{2}K_x^{\mathrm{D}}\right)=0.
\end{split}
\end{equation}

We now consider a strained honeycomb lattice  in which the hopping amplitude along the $y$ direction acquires a linear spatial dependence, $t_y=t(1+\alpha y)$.
As a consequence, the Dirac point is shifted from $\mathbf{K}^{\mathrm{D}}$ to $\mathbf{K}^{\mathrm{S}}=\mathbf{K}^{\mathrm{D}}+\boldsymbol{\Delta}$, while the Dirac-point condition remains
\begin{equation}\label{h12_2}
\begin{split}
h_{12}=t(1+\alpha y)+2t\cos\!\left(\frac{a}{2}K_x^{\mathrm{S}}\right)=0.
\end{split}
\end{equation}
Expanding Eq.~\eqref{h12_2} around $\mathbf{K}^{\mathrm{D}}$ with $\boldsymbol{\Delta}=(\Delta_x,0)$ and retaining terms to linear order in $\Delta_x$, we obtain
\begin{align*}
t(1+\alpha y)
+2t\Bigg[
\cos\!\left(\frac{ a}{2}K_x^{\mathrm{D}}\right)
-\frac{ a}{2}
\sin\!\left(\frac{ a}{2}K_x^{\mathrm{D}}\right)\Delta_x
\Bigg]
\simeq 0,
\end{align*}

where we have used the fact that the strain-induced shift $\Delta_x$ is small compared to $K_x^{\mathrm{D}}$.
Employing Eq.~\eqref{h12}, the shift of the Dirac point is readily obtained as
\[
\Delta_x=\frac{2\alpha y}{\sqrt{3}a}.
\]
This allows us to identify the strain-induced pseudo-gauge potential as
\begin{equation}\label{Pseuda_Gauge}
\begin{split}
\mathbf{A}=(A_x,0,0)=(\Delta_x,0,0)=\left(\frac{2\alpha y}{\sqrt{3}a},0,0\right).
\end{split}
\end{equation}

To determine the effective Fermi velocities, we expand the Hamiltonian around the shifted Dirac point by substituting $\mathbf{k}=\mathbf{K}^{\mathrm{S}}+\boldsymbol{q}$.
The off-diagonal matrix element then reads
\begin{align*}
h_{12}
&=t(1+\alpha y)
+2t\cos\!\left[\frac{a}{2}\left(K_x^{\mathrm{S}}+ q_x\right)\right]
e^{i\frac{\sqrt{3}a}{2}q_y}.
\end{align*}

Using the identities,
\begin{align*}
\cos\!\left(\frac{ a}{2}K_x^{\mathrm{S}}\right)&=-\frac{1+\alpha y}{2},\\
\sin\!\left(\frac{ a}{2}K_x^{\mathrm{S}}\right)&=-\frac{\sqrt{3-2\alpha y-(\alpha y)^2}}{2},
\end{align*}
which follow directly from Eq.~\eqref{h12_2}, and expanding to first order in $\boldsymbol{q}$, we arrive at
\begin{equation}\label{Pseuda_Gauge_qxqy}
\begin{split}
h_{12}\simeq
\frac{\sqrt{3}at}{2}(1-\frac{\alpha y}{3})q_x
-i\frac{\sqrt{3}at}{2}(1+\alpha y)q_y.
\end{split}
\end{equation}

From Eq.~\eqref{Pseuda_Gauge_qxqy}, the strain-induced anisotropic Fermi velocities are identified as
\begin{align*}
v_x&=\frac{\sqrt{3}at}{2}(1-\frac{\alpha y}{3}),\\
v_y&=\frac{\sqrt{3}at}{2}(1+\alpha y).
\end{align*}

\section{Analytical solution under OBC}\label{AppB}
We start from the coupled equations describing the two-component wave function
\((\psi_A, \psi_B)^T\):
\begin{equation}
\begin{bmatrix}
0 & q_x - \frac{\partial}{\partial y} + \tilde{\delta} \\
q_x + \frac{\partial}{\partial y} - \tilde{\delta} & 0
\end{bmatrix}
\begin{pmatrix}
\psi_A \\ \psi_B
\end{pmatrix}
= \varepsilon
\begin{pmatrix}
\psi_A \\ \psi_B
\end{pmatrix}.
\end{equation}
The component form of this equation reads
\begin{align}
\left( q_x - \frac{\partial}{\partial y} + \tilde{\delta} \right) \psi_B &= \varepsilon \psi_A, \label{eq:comp1}\\
\left( q_x + \frac{\partial}{\partial y} - \tilde{\delta} \right) \psi_A &= \varepsilon \psi_B. \label{eq:comp2}
\end{align}
For $\varepsilon \neq 0$, we can express $\psi_B$ from Eq.~\eqref{eq:comp2} as
\begin{equation}
\psi_B = \frac{1}{\varepsilon} \left( q_x + \frac{\partial}{\partial y} - \tilde{\delta} \right) \psi_A.
\end{equation}
Substituting this into Eq.~\eqref{eq:comp1} gives
\begin{equation}
\varepsilon \psi_A =
\left( q_x - \frac{\partial}{\partial y} + \tilde{\delta} \right)
\left[ \frac{1}{\varepsilon}
\left( q_x + \frac{\partial}{\partial y} - \tilde{\delta} \right) \psi_A \right].
\end{equation}
Multiplying both sides by $\varepsilon$, we obtain
\begin{equation}
\varepsilon^2 \psi_A =
\left( q_x - \frac{\partial}{\partial y} + \tilde{\delta} \right)
\left( q_x + \frac{\partial}{\partial y} - \tilde{\delta} \right) \psi_A.
\end{equation}
Evaluating the operator product yields
\begin{equation}
\left( q_x - \frac{\partial}{\partial y} + \tilde{\delta} \right)
\left( q_x + \frac{\partial}{\partial y} - \tilde{\delta} \right)
= q_x^2 - \tilde{\delta}^2 + 2\tilde{\delta} \frac{\partial}{\partial y} - \frac{\partial^2}{\partial y^2}.
\end{equation}
Hence, $\psi_A(y)$ satisfies the second-order differential equation
\begin{equation}
\frac{\partial^2 \psi_A}{\partial y^2}
- 2\tilde{\delta} \frac{\partial \psi_A}{\partial y}
- \left(q_x^2 - \tilde{\delta}^2 - \varepsilon^2\right) \psi_A = 0.
\end{equation}

With open boundary conditions $\psi_A(0)=0$ and $\psi_A(L)=0$ (where $L=N_y$),
the characteristic equation gives the decay parameter
\begin{equation}
m = \tilde{\delta} \pm \sqrt{q_x^2 - \varepsilon^2}.
\end{equation}
The general solution can thus be written as
\begin{equation}
\psi_A(y) = e^{\tilde{\delta} y} \left( A e^{\lambda y} + B e^{-\lambda y} \right),
\qquad \lambda = \sqrt{q_x^2 - \varepsilon^2}.
\end{equation}
Applying the boundary conditions, we find
\begin{align}
\psi_A(0) &= A + B = 0 \quad \Rightarrow \quad B = -A, \\
\psi_A(L) &= A e^{\tilde{\delta} L} (e^{\lambda L} - e^{-\lambda L}) = 0
\quad \Rightarrow \quad e^{\lambda L} - e^{-\lambda L} = 0.
\end{align}
The latter condition requires $\lambda L = i n \pi$ with integer $n$,
leading to the quantization condition
\begin{equation}
\varepsilon^2 = q_x^2 + \left( \frac{n\pi}{L} \right)^2.
\end{equation}

Therefore, the energy eigenvalues are
\begin{equation}
\varepsilon_n = \pm \sqrt{ q_x^2 + \left( \frac{n\pi}{L} \right)^2 },
\qquad n = 1, 2, 3, \dots
\end{equation}
and the corresponding eigenfunctions are
\begin{align}
\psi_A(y) &= C\, e^{\tilde{\delta} y}
\sin\!\left( \frac{n \pi y}{L} \right), \\
\psi_B(y) &=
\frac{C}{\varepsilon_n} e^{\tilde{\delta} y}
\left[
q_x \sin\!\left( \frac{n \pi y}{L} \right)
+ \frac{n \pi}{L} \cos\!\left( \frac{n \pi y}{L} \right)
\right],
\end{align}
where $C$ is a normalization constant.

We start from the equation:
\begin{equation}
\begin{bmatrix}
0 & q_x - \frac{\partial}{\partial y} + \tilde{\delta} \\
q_x + \frac{\partial}{\partial y} - \tilde{\delta} & 0
\end{bmatrix}
\begin{pmatrix}
\psi_A \\ \psi_B
\end{pmatrix}
= \varepsilon
\begin{pmatrix}
\psi_A \\ \psi_B
\end{pmatrix}.
\end{equation}

The component equations are:
\begin{align}
\left( q_x - \frac{\partial}{\partial y} + \tilde{\delta} \right) \psi_B &= \varepsilon \psi_A, \\
\left( q_x + \frac{\partial}{\partial y} - \tilde{\delta} \right) \psi_A &= \varepsilon \psi_B.
\end{align}

Expressing \(\psi_B\) from the second equation (for \(\varepsilon \neq 0\)):
\begin{equation}
\psi_B = \frac{1}{\varepsilon} \left( q_x + \frac{\partial}{\partial y} - \tilde{\delta} \right) \psi_A.
\end{equation}

Substituting into the first equation:
\begin{equation}
\varepsilon \psi_A = \left( q_x - \frac{\partial}{\partial y} + \tilde{\delta} \right) \left[ \frac{1}{\varepsilon} \left( q_x + \frac{\partial}{\partial y} - \tilde{\delta} \right) \psi_A \right].
\end{equation}

Multiplying both sides by \(\varepsilon\):
\begin{equation}
\varepsilon^2 \psi_A = \left( q_x - \frac{\partial}{\partial y} + \tilde{\delta} \right) \left( q_x + \frac{\partial}{\partial y} - \tilde{\delta} \right) \psi_A.
\end{equation}

Computing the operator product:
\begin{equation}
\left( q_x - \frac{\partial}{\partial y} + \tilde{\delta} \right) \left( q_x + \frac{\partial}{\partial y} - \tilde{\delta} \right) = q_x^2 - \tilde{\delta}^2 + 2\tilde{\delta} \frac{\partial}{\partial y} - \frac{\partial^2}{\partial y^2}.
\end{equation}

Thus, we obtain:
\begin{equation}
\frac{\partial^2 \psi_A}{\partial y^2} - 2\tilde{\delta} \frac{\partial \psi_A}{\partial y} - (q_x^2 - \tilde{\delta}^2 - \varepsilon^2) \psi_A = 0.
\end{equation}

Solving this with boundary conditions \(\psi_A(0) = 0\) and \(\psi_A(L) = 0\) (where \(L = N_y\)), the characteristic equation gives:
\begin{equation}
m = \tilde{\delta} \pm \sqrt{q_x^2 - \varepsilon^2}.
\end{equation}

The general solution is:
\begin{equation}
\psi_A(y) = e^{\tilde{\delta} y} \left( A e^{\lambda y} + B e^{-\lambda y} \right), \quad \lambda = \sqrt{q_x^2 - \varepsilon^2}.
\end{equation}

Applying boundary conditions:
\begin{align}
&\psi_A(0) = A + B = 0 \implies B = -A, \\
&\psi_A(L) = A e^{\tilde{\delta} L} (e^{\lambda L} - e^{-\lambda L}) = 0 \implies e^{\lambda L} - e^{-\lambda L} = 0.
\end{align}

This requires \(\lambda L = i n \pi\) for integer \(n\), so:
\begin{equation}
\varepsilon^2 = q_x^2 + \left( \frac{n \pi}{L} \right)^2.
\end{equation}

Thus, the eigenvalues are:
\begin{equation}
\varepsilon_n = \pm \sqrt{ q_x^2 + \left( \frac{n \pi}{L} \right)^2 }, \quad n = 1, 2, 3, \dots.
\end{equation}

The eigenstates are:
\begin{align}
\psi_A(y) &= C e^{\tilde{\delta} y} \sin\left( \frac{n \pi y}{L} \right), \\
\psi_B(y) &= \frac{C}{\varepsilon_n} e^{\tilde{\delta} y} \left[ q_x \sin\left( \frac{n \pi y}{L} \right) + \frac{n \pi}{L} \cos\left( \frac{n \pi y}{L} \right) \right],
\end{align}
where \(C\) is a normalization constant.

\section{Derivation of Eigenvalues and Eigenstates of Eq.~\eqref{hq_Bs_dy0}}\label{app_B}
From Eq.~\eqref{hq_Bs_dy0}, when $y$-direction is open, we have the eigenequation
\begin{widetext}
\begin{equation}\label{hq_Bs_dy}
\begin{split}
\left[
  \begin{array}{cc}
    0 & q_x-B_s y-\frac{\partial}{\partial y}+\tilde{\delta} \\
    q_x-B_s y+\frac{\partial}{\partial y}-\tilde{\delta} & 0 \\
  \end{array}
\right]
\left(
  \begin{array}{c}
    \psi_A \\
    \psi_B \\
  \end{array}
\right)
=\varepsilon
\left(
  \begin{array}{c}
    \psi_A \\
    \psi_B \\
  \end{array}
\right),
\end{split}
\end{equation}
\end{widetext}
which reduces to a single second-order ordinary differential equation
\begin{equation}
\left[-\frac{\partial^2}{\partial y^2}+2\tilde{\delta}\frac{\partial}{\partial y}+
(q_x-B_s y)^2+B_s-\tilde{\delta}^2\right]\psi_A(y)=
\varepsilon^2 \psi_A(y).
\label{eq:supp_start}
\end{equation}
By performing the gauge-like transformation $\psi_A(y)=e^{\tilde{\delta} y}\phi(y)$, the first derivative term is eliminated and the equation reduces to
\begin{equation}
\left[-\frac{\partial^2}{\partial y^2}+(q_x-B_sy)^2+B_s\right]\phi(y)
=\varepsilon^2\phi(y).
\end{equation}
Introducing the shifted and rescaled variable
\begin{equation}
\eta=\sqrt{B_s}\left(y-\frac{q_x}{B_s}\right),
\end{equation}
we obtain the standard harmonic oscillator form
\begin{equation}
\left[-\frac{\mathrm{d}^2}{\mathrm{d}\eta^2}+\eta^2\right]\phi(\eta)
=\frac{\varepsilon^2-B_s}{B_s}\,\phi(\eta).
\end{equation}
This equation is identical to the Schr\"{o}dinger equation of a one-dimensional quantum harmonic oscillator, whose eigenvalues and eigenfunctions are well known. We thus find
\begin{equation}
\varepsilon_n=\pm\sqrt{2q\hbar B_sn}, \quad n=0,1,2,\ldots
\end{equation}
and
\begin{widetext}
\begin{equation}
\psi_{A,n}(y)=\mathcal{N}_n\,e^{\tilde{\delta} y}
\exp\!\left[-\tfrac{1}{2}B_s\left(y-\tfrac{q_x}{B_s}\right)^2\right]\,
H_n\!\Big(\sqrt{B_s}\left(y-\tfrac{q_x}{B_s}\right)\Big),
\end{equation}
\end{widetext}
where $H_n$ is the $n$-th Hermite polynomial and $\mathcal{N}_n$ is the normalization factor.
In the presence of finite boundary conditions, such as $\psi_A(0)=\psi_A(N_y)=0$, the allowed values of $q_x$ are further restricted, leading to a quantization of the guiding center of the Landau-like wavefunctions.

\end{appendix}

\bigskip

\end{document}